\documentclass[onecolumn,useAMS,usenatbib]{mn2e}
\usepackage{graphicx}
\usepackage{amsmath}

\title
  {Rotating gravitational lenses: a kinematic approach}
\author[Walters \& Forbes]
  {S.J. Walters and L.K. Forbes \\
  School of Mathematics and Physics, University of Tasmania, P.O. Box 37, Hobart, 7001, Tasmania, Australia}
\date{Released 2014 Xxxxx XX}

\begin{document}

\maketitle

\begin{abstract}
This paper uses the Kerr geodesic equations for massless particles to derive an acceleration vector in both Boyer-Lindquist and Cartesian coordinates. As a special case, the Schwarzschild acceleration due to a non-rotating mass has a particularly simple and elegant form in Cartesian coordinates. Using forward integration, these equations are used to plot the caustic pattern due to a system consisting of a rotating point mass with a smaller non-rotating planet. Additionally, first and second order approximations to the paths are identified, which allows for fast approximations of paths, deflection angles and travel-time delays.
\end{abstract}


\section{Introduction}
	The deflection of light rays by massive objects results in magnification of distant light sources when a massive body passes close to the line joining light source and observer. For a simple point-source-point-lens system, the magnification over time (the "light curve") has a smooth symmetric form, whereas a binary lensing system may produce significant deviations from the simple light curve, although these are typically of short duration. A comparison of such light curves is shown in Fig. 1. These were generated using one of the models mentioned in \citet{wf}, by sampling the number of light rays passing through a narrow strip of the magnification map.

\begin{figure}
\vspace{1cm}
\caption{Typical light curves for point-source-point-lens model (left), and for a binary system (right). The horizontal axis "Time" corresponds to distance that the observer has travelled across the magnification map. Intensity is relative to the un-lensed intensity of the background star.}
\includegraphics[height=4cm]{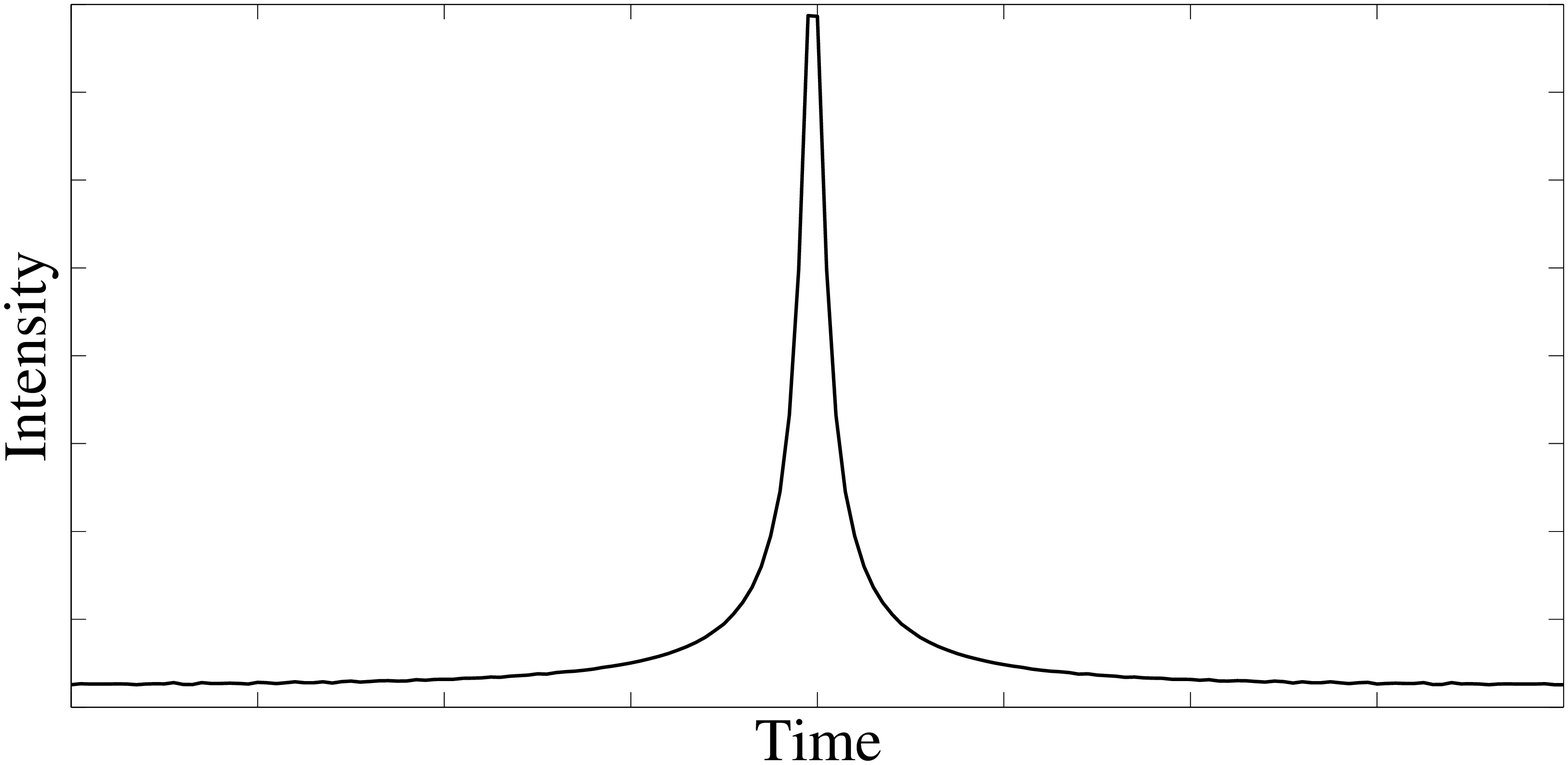} 
\includegraphics[height=4cm]{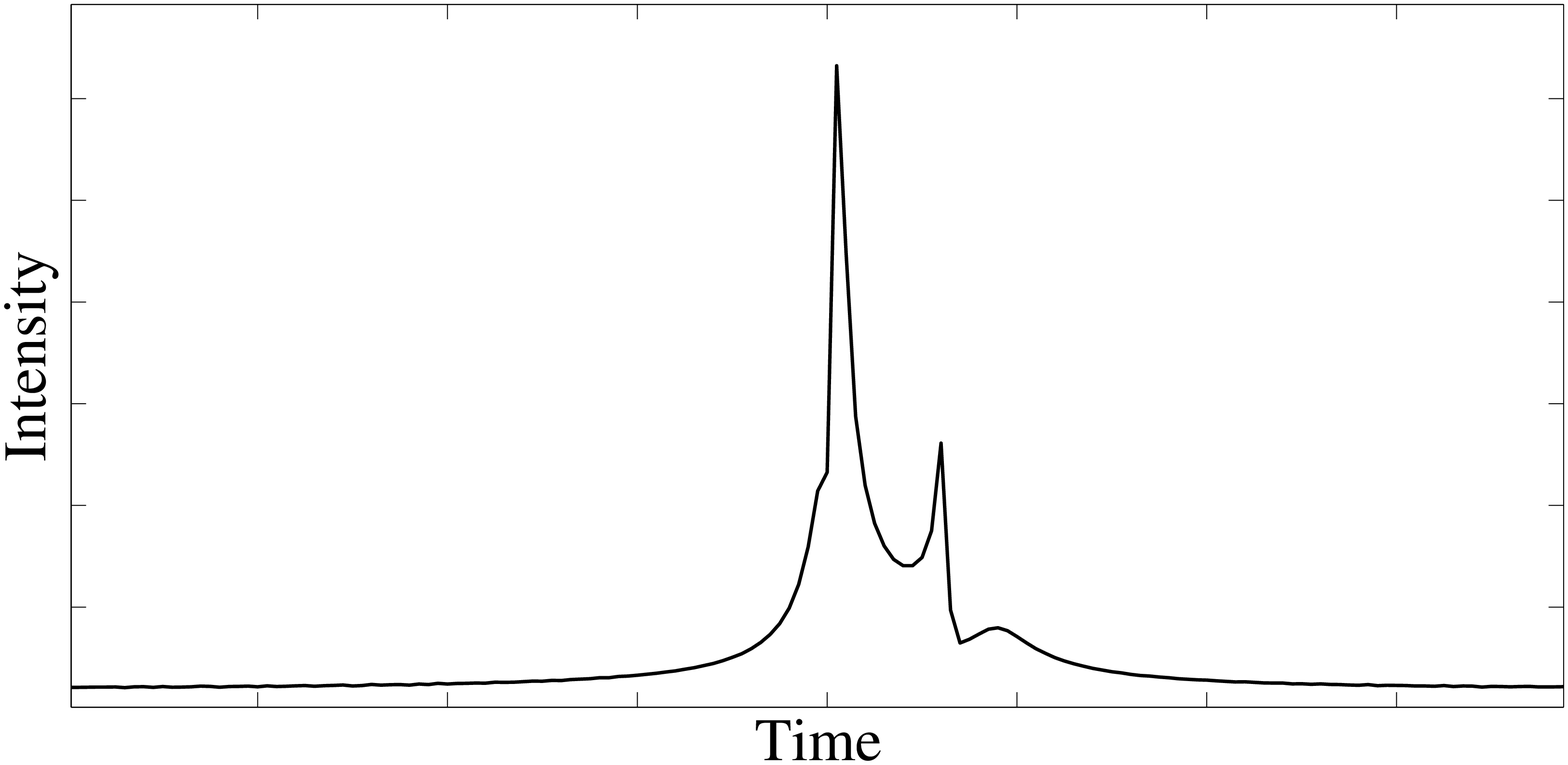}
\label{curves}
\end{figure}

 This phenomenon, known as gravitational lensing, is used by astrophysicists in identifying characteristics of the lensing object. Such an approach is useful in searching for dark matter, as suggested by \citet{pac}. The first exo-planet discovered using this approach was found in 2003 (see \citet{bond}), with several more discovered since that time. The presence of such planets in the lensing system can cause caustics in the magnification map. Such caustics are described by \citet{wam}. Various techniques can be used to model such caustic patterns. The simplest of these is to deflect the light ray as it crosses the lensing plane (this is the plane containing the lensing object, normal to the line joining the source and observer). The light path is then considered as two straight lines with an abrupt change of trajectory at the lensing plane. For a description of this method, see for example \citet{sch}. The amount of deflection in such a model is given by the Einstein deflection angle. As the deflection involved is very small (which means that the photon passes through areas of weak gravitational fields only), such a "first order" approach is a very accurate approximation.

Recently, \citet*{wfj} undertook a new approach, in which they used the Schwarzschild metric to derive kinematic type laws for the propagation of light rays through a lensing system. They found that the acceleration vectors thus derived gave results in close agreement with those obtained using the simpler model described above. Later, \citet{wf} considered a linearized approximation, in which the light rays were assumed to deflect only slightly from an otherwise straight-line path. They showed that their linearized equations were capable of an exact closed-form solution which agreed well with the fully non-linear simulations.
 In the present paper, the approach of \citet{wfj} is generalized to include the effects of relativistic frame dragging due to rotation of the lensing object, as described by the Kerr metric. A kinematic description is given in Section 2. It is found that converting to Cartesian co-ordinates simplifies the description of the light paths, by removing all acceleration terms at zeroth order. The non-rotating (Schwarzschild) case is examined in Section 3, and rotation effects, which become significant at second order in the Schwarzschild radius, are considered in Section 4. Application to delay of pulses in a binary pulsar model is presented in Section 5, and concluding remarks are given in Section 6.

\section{Light Rays in a Kerr System}

 The Kerr metric describes spacetime outside an uncharged point mass, rotating or otherwise. The Schwarzschild solution is contained as a special case wherein the mass has no angular momentum. Such a solution has spherical symmetry, whereas for a rotating body, the system is axi-symmetric only. For any light path other than one confined to the equatorial plane, a fully three dimensional description of the path is required. This is different than for the Schwarzschild case, where any path is confined to a plane, and can thus be treated as a two dimensional problem. We therefore begin with the Kerr metric given in Boyer-Lindquist coordinates (the conversion is described in Section \ref{acccomp} below), as written in Chandrasekhar's thorough mathematical treatment of black holes (\citet{cha})
 
\begin{eqnarray}
\mathrm{d} s^2&=&\frac{\Delta}{\rho^2}[\mathrm{d} t-(a\sin^2 \theta)\mathrm{d}\phi]^2-\frac{\sin^2 \theta}{\rho^2}[(r^2+a^2)\mathrm{d}\phi-a\mathrm{d}t]^2-\frac{\rho^2}{\Delta}(\mathrm{d}r)^2-\rho^2(\mathrm{d}\theta)^2.
\end{eqnarray}

From this metric, the equations of motion can be derived. In this paper, we are interested in the paths of light rays, so we consider the null geodesics $ds=0$ for a Kerr spacetime (\citet{cha}, pp. 346-7):

\begin{eqnarray}
\rho^4 \dot{r}^2&=&r^4+(a^2-L^2-Q)r^2+r_s r (Q+(L-a)^2)-a^2 Q \label{rdot} \\
\rho^4 \dot{\theta}^2&=&Q+a^2\cos^2{\theta}-L^2 \cot^2{\theta}\label{thetadot} \\
\rho^2 \dot{\phi}&=&\frac{1}{\Delta}(r_s a r+\frac{(\rho^2-r_s r)L}{\sin^2{\theta}})\label{phidot}\\
\rho^2 \dot{t}&=&\frac{1}{\Delta}((r^2+a^2)^2-r_s a r L).\label{tdot}
\end{eqnarray}

Here, the Schwarzschild radius is $r_{s}=2MG/c^{2}$, $t$ is the time coordinate in the reference frame of the mass, $a=J/Mc$ is the angular momentum term, and the dot indicates differentiation by a parameter, which we will call $\tau'$. The other symbols are defined as: $\rho^{2}=r^2 + a^2 \cos^{2}\theta$; $\Delta=r^2+a^2-r_s r$; $M$ is the mass of the body; and $J$ is the angular momentum of the body. We are using geometrized units, that is, $c=G=1$. Finally, $L$ and $Q$ are constants of the motion, related closely to the angular momentum of the particle. The first of these, $L$, comes from the first integral of the Euler-Lagrange equation for $\dot \phi$, and the second, $Q$, is Carter's constant, which is derived from the separation of the Hamilton-Jacobi equation for geodesic motion (\citet{cha}, p. 342).

\subsection{Acceleration Components} \label{acccomp}

Solving equations (\ref{rdot}) and (\ref{thetadot}) for $\dot{r}$ and $\dot{\theta}$ introduces square roots, for which the sign ($\pm$) is ambiguous (that is, either sign may be chosen). Additionally, we found that numerical integrators such as the Runge-Kutta method find singular solutions such as closed orbits when integrating these equations, and so do not always find the path of unbound photons. To remove these difficulties, we will take derivatives, producing acceleration components which have a simpler form than the first derivatives. As the parameterisation is arbitrary, for simplicity we first re-parameterise in order to remove the $\rho^2$ terms at the beginning of each equation. We choose a parameter $\tau$ such that $r^2 \frac{d}{d\tau}=\rho^2\frac{d}{d\tau'}$. This has the result that each instance of $\rho$ on the left of the geodesic equations above becomes $r$. Re-using the dot-notation for $d/d\tau$ and differentiating gives the following equations:
\begin{eqnarray}
\ddot{r}&=&\frac{L^2+Q-a^2}{r^3}-\frac{3 r_s}{2 r^4}(Q+(L-a)^2)+\frac{2 a^2 Q}{r^5} \label{rdd} \\
\ddot{\theta}&=&\frac{\cos{\theta}}{\sin^3\theta}(L^2-a^2\sin^4\theta)-\frac{2 \dot{r}\dot{\theta}}{r} \label{thetadd} \\
\ddot{\phi}&=&\frac{a \dot{r}}{r^2\Delta^2}(r_s a^2 - r_s r^2 + a L (2 r -r_s)) - 
  \frac{ 2 L \cos{\theta}}{r^2\sin^3{\theta}} \dot\theta - \frac{2\dot{r}\dot\phi}{r}
\label{phidd}
\end{eqnarray}

In order to describe the path of a particle through a system consisting of more than a single body at the origin, it is convenient to express the acceleration components in Cartesian co-ordinates. The conversion is given by the following substitutions (\citet{cha}, pp. 306-7):
\begin{eqnarray}
x&=&(r \cos\widetilde{\varphi}+a \sin\widetilde{\varphi})\sin\theta \nonumber \\
y&=&(r \sin\widetilde{\varphi}-a \cos\widetilde{\varphi})\sin\theta \nonumber \\
z&=&r \cos\theta
\label{xyz}\end{eqnarray}
where $\dot{\widetilde{\varphi}}=\dot\phi-a \dot{r}/ \Delta$. These equations provide an implicit definition of $r$ as:
\begin{eqnarray}
r^4-r^2(x^2+y^2+z^2-a^2)-a^2 z^2=0 \nonumber
\end{eqnarray}

Notice that if there is no rotation, that is, $a=0$, then this degenerates to a conversion from spherical co-ordinates, as expected. Differentiating the first equation in (\ref{xyz}) twice gives the following expression for $\ddot x$:

\begin{eqnarray}
\ddot x=\frac{1}{a^2+r^2}\biggl[(x+\frac{r_s a y}{\Delta})(r\ddot{r}+\frac{a^2-r^2}{a^2+r^2}\dot r^2)+ r \dot r \big(\dot x+\frac{r_s a \dot y}{\Delta}-\frac{r_s a (2 r-r_s)y \dot r}{\Delta^2}\big)\biggl]-\dot y \dot\phi-y \ddot\phi -\frac{x
\dot\theta^2}{\sin^2\theta}+\frac{\dot x \dot\theta +x \ddot\theta}{\tan \theta}.
\label{xdd}
\end{eqnarray}

In this equation, $\dot r$, $\dot \theta$ and $\dot \phi$ are obtained from the conversion equations (\ref{xyz}) by differentiation. A similar approach for $y$ and $z$ will give expressions for $\ddot y$ and $\ddot z$ respectively. Substituting in equations (\ref{rdd})-(\ref{phidd}) for $\ddot r$, $\ddot \theta$ and $\ddot \phi$ and simplifying leads to a system of the form
\begin{eqnarray}
\ddot x=\frac{-3 r_s x (L^2+Q)}{2 r^5}+a F_x(x,y,z,\dot x,\dot y,\dot z) \nonumber \\
\ddot y=\frac{-3 r_s y (L^2+Q)}{2 r^5}+a F_y(x,y,z,\dot x,\dot y,\dot z) \nonumber \\
\ddot z=\frac{-3 r_s z (L^2+Q)}{2 r^5}+a F_z(x,y,z,\dot x,\dot y,\dot z)
\label{xyzdd2}
\end{eqnarray}

The constant $a$ has a valid range from $-r_s/2$ to $r_s/2$. It is therefore reasonable to say that the angular momentum term $a$ is of the same order of magnitude as the Schwarzschild radius $r_s$. It may then be said that because the functions $F_x, F_y$ and $F_z$ are of order $r_s$, the first term in each of the equations in (\ref{xyzdd2}) is of first order, and the remainder is second order and higher. The full acceleration components in equation (\ref{xyzdd2}) are given in the appendix.

\section{Schwarzschild Acceleration in Cartesian Co-ordinates} \label{secschw}
We can see that for the non-rotating (Schwarzschild) case, that is, $a=0$, we obtain the elegant result:

\begin{eqnarray}
\mathbf{\ddot{r}}=\frac{-3 r_s (L^2+Q)}{2 r^5}\mathbf{r}
\label{schw1}
\end{eqnarray}
where $\mathbf{r}=[x,y,z]$ is the position vector, and $r=\lvert \lvert \mathbf{r} \rvert \rvert=\sqrt{x^2+y^2+z^2}$ is its Euclidean distance from the origin. From the non-rotating ($a=0$) versions of equations (\ref{thetadot}) and (\ref{phidot}) and the conversion equations (\ref{xyz}), we can write 
\begin{eqnarray}
L&=&x \dot y-y \dot x \nonumber \\
Q&=&(x \dot z-z \dot x)^2+(z \dot y-y \dot z)^2.
\label{LQ}
\end{eqnarray}

We can now say that $L^2+Q$ is the square of the impact parameter, which is the perpendicular distance of the initial (straight-line) path of the photon from the point lens. Equation (\ref{schw1}) is presented in a form similar to the standard Newtonian gravitational equation

\begin{eqnarray}
\mathbf{\ddot{r}}=\frac{-r_s}{2 r^3}\mathbf{r}. \nonumber
\end{eqnarray}
However, it should be noted that the parameter in equation (\ref{schw1}) differs in that it includes the time dilation factor, that is, $\dot t=r/(r-r_s) $. It will be helpful to explore the Schwarzschild solution in this coordinate system before continuing on to the more general Kerr solution. Expanding equation (\ref{schw1}) into the three components gives the equality:

\begin{eqnarray}
\frac{\ddot x}{x}=\frac{\ddot y}{y}=\frac{\ddot z}{z}, \nonumber
\end{eqnarray}
which can be integrated to give the angular momentum conservation equations, analogously with classical mechanics:
\begin{eqnarray}
x \dot y-y \dot x&=&L_z \nonumber \\
x \dot z-z \dot x&=&L_y \nonumber \\
y \dot z-z \dot y&=&L_x. \nonumber
\end{eqnarray}
In these equations the constants $L_x$, $L_y$ and $L_z$ are the three components of angular momentum. From equation (\ref{phidot}), we can identify $L$ with $L_z$. Taking the inner product of equation (\ref{schw1}) with $\bf \dot r$ and integrating gives

\begin{eqnarray}
\lvert \lvert \mathbf{\dot r} \rvert \rvert^2=1+\frac{r_s(L^2+Q)}{r^3}, \nonumber
\end{eqnarray}
after the integration constant has been determined by the boundary condition $\lvert \lvert \mathbf{\dot r} \rvert \rvert \rightarrow 1$ as $r \rightarrow \infty$. Further use of the identity (\ref{LQ}) enables this to be expressed in the final form
\begin{eqnarray}
(x \dot x+y \dot y+z \dot z)^2=x^2+y^2+z^2-(L_x^2+L_y^2+L_z^2)+\frac{r_s}{\sqrt{(x^2+y^2+z^2)}}(L_x^2+L_y^2+L_z^2).
\label{schw3}
\end{eqnarray}
Equation (\ref{schw3}) permits us to identify $Q$ with $L_x^2+L_y^2$  and we arrive back at the non-rotating version of (\ref{rdd}). We have identified $Q$ and $L$ in the non-rotating case with the angular momentum of the particle. In the rotating case, we will see that while there are conserved quantities, $Q$ and $L$, they are not identical with $L_x^2+L_y^2$ and $L_z$ above. Due to the spherical symmetry of the Schwarzschild system, $L$ and $Q$ only appear in the form $L^2+Q$. For readability in the Schwarzschild analysis to follow, it is convenient to introduce the non-negative constant $K=L^2+Q$.

\subsection{Linearized Schwarzschild Expansion} \label{secschw1}

We can approximate the path taken by photons in the Schwarzschild system, using the expansions:

\begin{eqnarray}
x&=&X_0+r_s X_1+r_s^2 X_2+O(r_s^3) \nonumber \\
y&=&Y_0+r_s Y_1+r_s^2 Y_2+O(r_s^3) \nonumber \\
z&=&Z_0+r_s Z_1+r_s^2 Z_2+O(r_s^3)
\label{xyz1}
\end{eqnarray}
where $r_s$ is considered small, relative to the distance of closest approach. Matching terms of corresponding order in $r_s$ will give the zeroth, first and second order solutions.  Differentiating the first equation in (\ref{xyz1}) twice and equating with the $x$-component of equation (\ref{schw1}) yields:
\begin{eqnarray}
\ddot X_0+r_s \ddot X_1=\frac{-3 r_s x K}{2 r^5}+O(r_s^2), \nonumber
\end{eqnarray}
where instances of $x$, $y$ and $z$ in the right side must also be expanded. Matching up the zeroth-order terms gives $\ddot X_0=0$ (and similarly $\ddot Y_0=0$ and $\ddot Z_0=0$). Integrating twice gives us the zeroth order solution

\begin{eqnarray}
X_0&=&C_1 \tau+C_2 \nonumber \\
Y_0&=&C_3 \tau+C_4 \nonumber \\
Z_0&=&C_5 \tau+C_6
\label{xyz0}
\end{eqnarray}
for some constants of integration $C_1$ to $C_6$. As is expected, this solution describes a straight line. In order to solve the first-order and second-order equations, it will be necessary to expand $r$ and $K=L^2+Q$ in powers of $r_s$ using equation (\ref{xyz1}). We write $r=R_0+r_s R_1+r_s^2 R_2+O(r_s^3)$ and then the zeroth order term for $r$ is given by
\begin{eqnarray}
R_0^2&=&X_0^2+Y_0^2+Z_0^2 \nonumber \\
&=&A \tau^2+2 B \tau+C \nonumber
\end{eqnarray}
where we have introduced three constants for readability:
\begin{eqnarray}
A&=&C_1^2+C_3^2+C_5^2 \nonumber \\
B&=&C_1 C_2+C_3 C_4+C_5 C_6 \nonumber \\
C&=&C_2^2+C_4^2+C_6^2. \label{abc}
\end{eqnarray}
However, we note that in the zeroth order solution, the speed of the photon ($=\sqrt{C_1^2+C_3^2+C_5^2}$) is $1$, so that $A=1$. The zeroth order term for $K$ is
\begin{eqnarray}
K_0&=&(X_0 \dot Y_0-Y_0 \dot X_0)^2+(X_0 \dot Z_0-Z_0 \dot X_0)^2+(Z_0 \dot Y_0-Y_0 \dot Z_0)^2 \nonumber \\
&=&C-B^2 \nonumber
\end{eqnarray}

Terms of first order in the small parameter $r_s$ are now equated and we obtain
\begin{eqnarray}
\ddot X_1=\frac{-3 X_0 K_0}{2 R_0^5}. \nonumber
\end{eqnarray}
We can now use the substitution $\tau+B=\sqrt{K_0} \tan \gamma$ and integrate twice. This gives the first-order corrections to the light paths
\begin{eqnarray}
X_1&=&\frac{X_0}{2 R_0}-\frac{R_0}{K_0}(C_2-B C_1)+C_{11} \tau + C_{21} \nonumber \\
Y_1&=&\frac{Y_0}{2 R_0}-\frac{R_0}{K_0}(C_4-B C_3)+C_{31} \tau + C_{41} \nonumber \\
Z_1&=&\frac{Z_0}{2 R_0}-\frac{R_0}{K_0}(C_6-B C_5)+C_{51} \tau + C_{61}
\label{XYZ1}
\end{eqnarray}
Consequently, the first-order velocity components are:
\begin{eqnarray}
\dot X_1&=&\frac{C_1}{2 R_0}-\frac{X_0 (\tau+B)}{2 R_0^3}-\frac{\tau+B}{R_0 K_0}(C_2-B C_1)+C_{11} \nonumber \\
\dot Y_1&=&\frac{C_3}{2 R_0}-\frac{Y_0 (\tau+B)}{2 R_0^3}-\frac{\tau+B}{R_0 K_0}(C_4-B C_3)+C_{31} \nonumber \\
\dot Z_1&=&\frac{C_5}{2 R_0}-\frac{Z_0 (\tau+B)}{2 R_0^3}-\frac{\tau+B}{R_0 K_0}(C_6-B C_5)+C_{51}
\label{XYZ1d}
\end{eqnarray}

Choosing the initial position for the light ray gives us the three constants $C_2, C_4, C_6$. We then specify the initial angle of the ray by choosing two of $\dot x$, $\dot y$ and $\dot z$, and the third of these can be identified using the geodesic equations (\ref{rdot}), (\ref{thetadot}), and (\ref{phidot}) to determine the speed: 
\begin{eqnarray}
\dot x^2+\dot y^2+\dot z^2=\dot r^2+r^2 \sin^2 \theta \dot \phi^2+r^2 \dot \theta^2=1+r_s K/r^3. \label{speed}
\end{eqnarray}
This gives the constants $C_1$, $C_3$ and $C_5$. We can then solve for $C_{11}$ to $C_{61}$ in the same way using the equations in (\ref{XYZ1}) and (\ref{XYZ1d}) and the speed equation (\ref{speed}). We now have complete path equations for the first order approximation. Converting the velocity given by equations (\ref{rdot})-(\ref{phidot}) (with $a=0$) to Cartesian co-ordinates gives a constraint on the constants of integration which will be useful later:
\begin{eqnarray}
C_1 C_{11}+C_3 C_{31}+C_5 C_{51}=0. \nonumber
\end{eqnarray}

\subsection{Application: Magnification map - binary system}
We are now in a position to determine the caustic map due to photons travelling through a system consisting of one or more non-rotating masses, either by tracing their paths using forward integration of equation (\ref{schw1}) or by solving the first order equations as above. In either case, we determine the initial conditions for each light ray. With the lens placed at the origin, we place the source of the rays on the $x$-axis, at $(x_{source},0,0)$. This source is taken to emit light isotropically, so the rays are spread evenly over the azimuthal angle $\phi_s$ from $0$ to $2 \pi$ and the inclination angle $\theta_s$ from $-\pi/2$ to $\pi/2$. To save on computational time, we will only include the small subset of these rays that will pass near to the planet. For each ray, $dy/dx=\tan \phi_s$ and $dz/dx=\tan \theta_s \sec \phi_s$. This gives us five of the six initial values for the ray. The speed of the photon is determined by the speed equation (\ref{speed}), substituting in the five chosen initial values in the equations for $K$ and for the speed (\ref{speed}) to obtain an equation for $\dot x$:

\begin{eqnarray}
\dot x^2=\left(1+ \bigg[\left( \frac{dy}{dx}\right)^2+\left(\frac{dz}{dx}\right)^2\bigg]\bigg[ 1-\frac{r_s}{ \left|x_{source}\right| }\bigg] \right)^{-1}  \nonumber
\end{eqnarray}

We can then say that $\dot y=\dot x (dy/dx)$ and $\dot z=\dot x (dz/dx)$. Having the six initial values ($x,y,z,\dot x,\dot y,\dot z$), forward integration can now be used to solve the system of six first order equations
\begin{equation}
\frac{d}{d\tau}
\left[\begin{array}{c}
x\\
y\\
z\\
\dot x\\
\dot y\\
\dot z\\
\end{array}
\right]=\left[\begin{array}{c}
\dot x\\
\dot y\\
\dot z\\
\ddot x\\
\ddot y\\
\ddot z
\nonumber
\end{array}\right]
\end{equation}
where the acceleration components $\ddot x, \ddot y$ and $\ddot z$ are given by equation (\ref{schw1}). The integration is stopped once the value of $x$ corresponds to that of the observer, and a point is then plotted at ($y,z$).

Unsurprisingly, tracing such paths through a system consisting of a central mass and a single planet produces the same diamond caustic pattern, similar to that seen in the top part of Fig. \ref{kerr1} in Section 4, which was described by \citet{wam}, and was also plotted previously using 2-dimensional polar co-ordinates in \citet{wfj} and \citet{wf}. Interestingly the computations were slightly quicker with this new Cartesian system, as it was not necessary to rotate each ray into the $x,y$ or $r,\phi$ plane, and also because the zeroth order terms of $\ddot x$, $\ddot y$ and $\ddot z$ in equation (\ref{schw1}) are now all zero (whereas those of $\ddot r$ and $\ddot \phi$ are not). This leaves only small acceleration terms which the numerical integration routine can process more rapidly. However, any simple forward integration method, including this one, is still slower than more elaborate methods, such as the semi-analytical method of \citet{dex} or the numerical methods of \citet{rb} which increase computational efficiency through the use of elliptic integrals and clever changes of variables.

Similarly, to obtain the first order solution (\ref{XYZ1}) calculated above, we again use the five given initial conditions and the speed equation to give ($x,y,z,\dot x,\dot y,\dot z$). These are used with the zeroth order equations (\ref{xyz0}) and their derivatives to derive the constants $C_1-C_6$ and then with the first order equations (\ref{XYZ1}) and (\ref{XYZ1d}) to derive $C_{11}-C_{61}$. The position of the ray ($y,z$) at the observer's plane can then be directly calculated from the zeroth and first order equations, by solving for $\tau$ at the value of $x$ corresponding to the observer, and then applying that value of $\tau$ to the equations for $y$ and $z$. This gives a caustic map indistinguishable from that obtained using forward integration, but in a much shorter time, approaching the speed of the thin lens formula $\delta=r_s/b$.

\subsection{Application: Total Deflection Angle - first order approximation} \label{sectiondeflect1}
\begin{figure}
\vspace{1cm}
\caption{Approximating deflection and delay to the light path near a massive object, located at the origin. For ease of calculations, the light path is chosen so that the ray is horizontal as it crosses the $y$-axis. For a non-rotating mass, the path is left-right symmetric.}
\includegraphics[width=15cm]{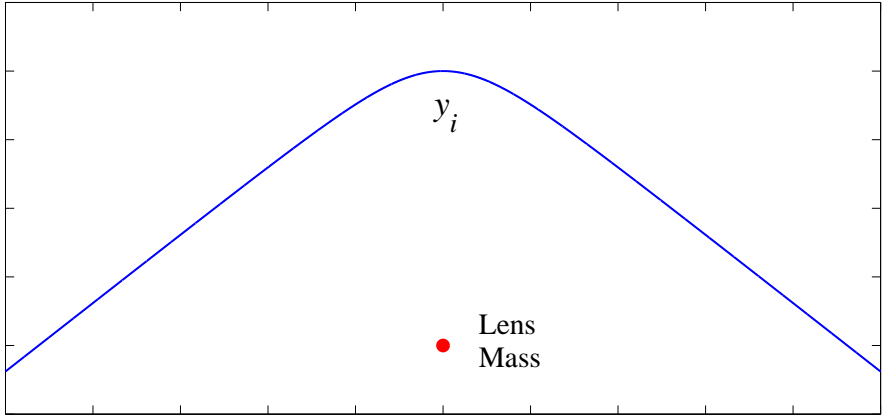} \\
\label{fig2}
\end{figure}

The well known total deflection for a light ray passing near to a spherically symmetric mass can now easily be estimated to first order in $r_s$. Due to the spherical symmetry of the space-time around the non-rotating mass, we can choose a ray confined to the equatorial plane, without loss of generality. At $\tau=0$, let the ray cross the $y$-axis parallel to the $x$-axis, at some value $y_i$, as shown in Fig. \ref{fig2}. Solving for the speed of the particle at $\tau=0$, (where $\dot y=0$), it can be seen that $\dot x^2=1+r_s K/r^3$. Also, at that point, $x=0$ and $y=y_i$. It is straightforward to solve for the zeroth-order constants and obtain $C_1=1$, $C_2=0$, $C_3=0$, $C_4=y_i$. The first-order constants can then be calculated to give $C_{11}=0$, $C_{21}=0$, $C_{31}=0$, $C_{41}=1/2$. Having the full first-order path equations, the total deflection is given by the difference in $\arctan (\dot y/\dot x)$ as $\tau \rightarrow \infty$ and $\arctan (\dot y/\dot x)$ as $\tau \rightarrow -\infty$. This gives the result $2r_s/y_i+O(r_s^2)$, which is consistent with the well known Einstein deflection. In this case, $y_i$ is the point of closest approach (often referred to as $r_0$), and also the zeroth-order approximation to the impact parameter, often referred to as $b$. Thus to first-order in $r_s$, $2r_s/y_i=2r_s/b$.

\subsection{Application: Travel Time Delay - first order approximation} \label{delaysec1}
Using the first-order equations again, it is a simple matter to compute the travel-time for a photon from any initial point and time ($x_i, y_i, \tau_i$) to any other point and time ($x_f, y_f, \tau_f$). For ease of computation, and without loss of generality, we may use the same arrangement, and therefore the same constants as described in the angle calculation illustrated in Fig. \ref{fig2} (Section \ref{sectiondeflect1}). In order to measure the travel-time to a point of given radius $r_f$, we solve for $\tau_f$ by means of the path equations with the constraint $x_f^2+y_f^2=r_f^2$. This will simplify the calculation of the travel time delay for a light ray passing close to the sun. This delay has been calculated to first order previously, and will serve as a check on this new method. We note that $y_i$ is the closest approach to the sun, which is usually designated $r_0$, Then, at the final point, $\tau=\tau_f$, so $X_0=\tau_f$ and $Y_0=r_0$, so that at that point, $R_0=\sqrt{\tau_f^2+r_0^2}$. The first order terms are
\begin{eqnarray}
X_1&=&\frac{\tau_f}{2\sqrt{\tau_f^2+r_0^2}} \nonumber \\
Y_1&=&\frac{r_0}{2\sqrt{\tau_f^2+r_0^2}}-\frac{\sqrt{\tau_f^2+r_0^2}}{r_0}+\frac{1}{2}. \nonumber
\end{eqnarray}
To obtain the first-order delay term, we solve for $\tau_f$, and then convert to co-ordinate time $t$ by equation (\ref{tdot}), giving
\begin{eqnarray}
r_f^2&=&x_f^2+y_f^2 \nonumber \\
&=&\bigg[\tau_f+r_s\frac{\tau_f}{2\sqrt{\tau_f^2+r_0^2}}\bigg]^2+\bigg[r_0+r_s\bigg(\frac{r_0}{2\sqrt{\tau_f^2+r_0^2}}-\frac{\sqrt{\tau_f^2+r_0^2}}{r_0}+\frac{1}{2}\bigg)\bigg]^2+O(r_s^2) \nonumber \\
&=&\tau_f^2+r_0^2+r_s(r_0-\sqrt{\tau_f^2+r_0^2})+O(r_s^2). \label{delaycalc1}
\end{eqnarray}
This is a quadratic equation in $\sqrt{\tau_f^2+r_0^2}$. After solving, we see that 
\begin{eqnarray}
\tau_f=\pm\sqrt{r_f^2-r_0^2}(1+\frac{r_s}{2(r_f+r_0)})+O(r_s^2).\nonumber
\end{eqnarray}
Solving equation (\ref{tdot}) to first order and integrating gives $t=\tau+r_s \ln((\tau+R_0)/r_0)+O(r_s^2)$, the constant of integration being determined by letting $t=0$ when $\tau=0$. Substituting this into equation (\ref{delaycalc1}) gives the total travel time
\begin{eqnarray}
t_f&=&\pm\bigg(\sqrt{r_f^2-r_0^2}+\frac{r_s}{2}\sqrt{\frac{r_f-r_0}{r_f+r_0}}+r_s \ln{\frac{r_f+\sqrt{r_f^2-r_0^2}}{r_0}}\bigg)+O(r_s)^2. \label{traveltime1}
\end{eqnarray}
The first term on the right hand side of equation (\ref{traveltime1}) is the straight-line time, and the rest constitutes the delay. This delay is in complete agreement with the well known first order delay (for example, see \citet{wei} p.202).

\subsection{Second Order Schwarzschild Expansion} \label{secschw2}

Frame dragging effects due to rotation do not occur at first order, so it will be necessary to consider the Kerr metric equations at second order. Before doing so, it will be worth identifying the second order expansion of the Schwarzschild system. The advantage of this approach is that we can follow the same procedure as above while dealing with fewer terms than in the full rotational model.

The second-order terms $X_2$, $Y_2$ and $Z_2$ in the expansion (\ref{xyz1}) are now considered. First it is necessary to expand $r=\sqrt{x^2+y^2+z^2}$ and the constant $K$ to first order in $r_s$, that is, $r=R_0+r_s R_1+O(r_s^2)$ and $K=K_0+r_s K_1+O(r_s^2)$. From Section \ref{secschw1}, it is straightforward to establish that 

\begin{eqnarray}
R_1&=&-\frac{1}{2}+\frac{1}{R_0}(B_{R} \tau+C_{R}) \nonumber \\
K_1&=&2(C_R-B_R B) \nonumber
\label{r1eq2}
\end{eqnarray}
where we have introduced two more constants, $B_R$ and $C_R$ for readability. These are named according to their similarity with the constants $B$ and $C$ in equations (\ref{abc}). They are 

\begin{eqnarray}
B_{R}&=&C_1 C_{21}+C_2 C_{11}+C_3 C_{41}+C_4 C_{31}+C_5 C_{61}+C_6 C_{51} \nonumber \\
C_{R}&=&C_2 C_{21}+C_4 C_{41}+C_6 C_{61}. \nonumber
\end{eqnarray}
In a manner similar to the first order expansion of section \ref{secschw1}, we can now expand $\ddot x$ to second order, and equation (\ref{schw1}) yields

\begin{eqnarray}
\ddot X_0+r_s \ddot X_1+r_s^2 \ddot X_2=\frac{-3 r_s (X_0+r_s X_1)(K_0+r_s K_1)}{2 (R_0+r_s R_1)^5}+O(r_s^3). \nonumber
\end{eqnarray}
Expanding and matching terms with coefficient $r_s^2$ gives:
\begin{eqnarray}
\ddot X_2&=&\frac{-3 K_0}{2 R_0^5}\big(X_1-5 X_0 R_1/R_0+ K_1 X_0/K_0\big) \nonumber\\
&=&\frac{-3 K_0}{2 R_0^5}\bigg(3 \frac{X_0}{R_0}-\frac{R_0}{K_0}(C_2-B C_1)+C_{11} \tau + C_{21} -5\frac{X_0}{R_0^2} (B_{R} \tau+C_{R})+\frac{K_1}{K_0}X_0\bigg). \nonumber
\end{eqnarray}
Integrating twice gives the equation for $X_2$:

\begin{eqnarray}
X_2 = \frac{X_0}{R_0}F_1 +\frac{C_1}{\sqrt{K_0}}F_2+\frac{C_1 B-C_2}{K_0}F_3 -\frac{C_{21} - C_{11} B}{K_0}R_0+\frac{C_{11}\tau+C_{21}}{2R_0}+C_{12} \tau+C_{22}.
\label{x2int}
\end{eqnarray}
The intermediary functions $F_1$, $F_2$ and $F_3$ are given by:

\begin{eqnarray}
F_1&=&\frac{9}{16 R_0}-\frac{B_{R} \tau+C_{R}}{2 R_0^2} \nonumber\\
F_2&=&\frac{B_{R} R_0}{\sqrt{K_0}}+\frac{9}{16} \arctan{\frac{\tau + B}{\sqrt{K_0}}} \nonumber\\
F_3&=&2 R_0\frac{B_{R} B-C_{R}}{K_0}+\frac{B_{R} \tau+C_{R}}{R_0}+\frac{15}{16}\frac{\tau+B}{\sqrt{K_0}} \arctan{\frac{\tau + B}{\sqrt{K_0}}} \nonumber
\end{eqnarray}
Due to the spherical symmetry of the Schwarzschild space-time, the equations for $Y_2$ and $Z_2$ have a similar form:

\begin{eqnarray}
Y_2&=&\frac{Y_0}{R_0}F_1 +\frac{C_3}{\sqrt{K_0}}F_2+\frac{C_3 B-C_4}{K_0}F_3 -\frac{C_{41} - C_{31} B}{K_0}R_0+\frac{C_{31}\tau+C_{41}}{2R_0}+C_{32} \tau+C_{42} \nonumber\\
Z_2&=&\frac{Z_0}{R_0}F_1 +\frac{C_5}{\sqrt{K_0}}F_2+\frac{C_5 B-C_6}{K_0}F_3 -\frac{C_{61} - C_{51} B}{K_0}R_0+\frac{C_{51}\tau+C_{61}}{2R_0}+C_{52} \tau+C_{62}
\label{yz2int}
\end{eqnarray}

As in Section \ref{secschw1}, we can identify the constants, $C_{12}$, $C_{32}$ and $C_{52}$ by solving for $\dot X_2$, $\dot Y_2$ and $\dot Z_2$ at $\tau=0$, and likewise to determine $C_{22}$, $C_{42}$ and $C_{62}$ we solve for $X_2$, $Y_2$ and $Z_2$ at $\tau=0$. We can now compare the paths taken by light rays as calculated using the following three methods:

(i) forward integration of equation (\ref{schw1});

(ii) zeroth-order path equations (\ref{xyz0}) with first-order corrections (\ref{XYZ1}); and

(iii)  zeroth-order path equations (\ref{xyz0}) with first-order corrections (\ref{XYZ1}) and second-order corrections (\ref{x2int}) and (\ref{yz2int}).

\begin{figure}
\vspace{1cm}
\caption{Comparison of first and second order path approximations against numerical integration of the full acceleration vector. Rays originate at (-10,0) and are deflected by the mass at (0,0) of Schwarzschild radius $0.2$. Five different initial trajectories are chosen, each of which is computed using the three different methods. In each case, the deflection is greatest with the forward integration of the acceleration vector, and least with the first order path equations.}
\includegraphics[width=16cm]{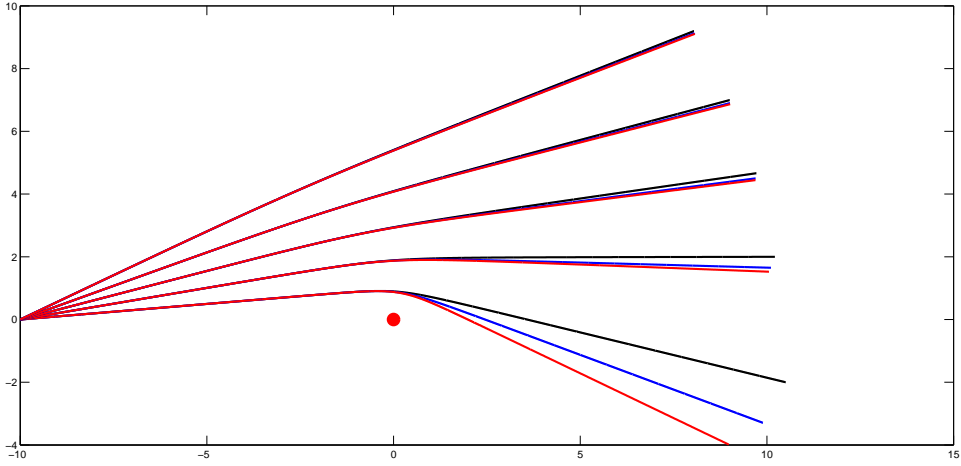}
\label{schw_cf}
\end{figure}

As the paths within the Schwarzschild system are contained within a plane, we can compare our different solutions in two dimensions without loss of generality. This is illustrated in Fig. \ref{schw_cf} with $r_s=0.2$, where the three different methods have been applied to five rays originating from (0,-10) with different starting angles, and each being deflected by the mass at the origin. The three methods agree well in the weak gravity regime at the top of the diagram, and the second order solution does not diverge much from the exact solution until the deflection becomes quite large, that is for rays passing close to the mass. In each case, the second order paths approximate the paths obtained by forward integration more closely than do the first order paths. In particular, for selected points, if we call the difference in deflection angle between the numerically calculated path and the first-order path $\delta_1$, and the difference in deflection angle between the numerically calculated path and the second-order path $\delta_2$, we find that $\delta_2 \approx \delta_1^2$. That is, the errors are found to behave proportional to $r_s$ and $r_s^2$, as expected.

\subsection{Application: Total Deflection angle - second order approximation}
Following the earlier procedure for the first order approximation of the total deflection angle in Section \ref{sectiondeflect1}, we can now easily determine the second order correction. Of the second order constants, only $C_{32}$ will appear in this calculation, and by noting that $\dot y=0$ at $\tau=0$, its value is found to be $C_{32}=0$. The deflection angle is again given by the difference in $\arctan (\dot y/\dot x)$ as $\tau \rightarrow \infty$ and $ \arctan (\dot y/\dot x)$ as $\tau \rightarrow -\infty$. In the system described in Section \ref{sectiondeflect1}, and represented in Fig. \ref{fig2}, this is approximated by
\begin{eqnarray}
\Delta\Phi&=&2 \frac{\dot y}{\dot x}\bigg|_{\tau \rightarrow \infty}=2\frac{\dot Y_0+r_s\dot Y_1+r_s^2\dot Y_2}{\dot X_0+r_s\dot X_1+r_s^2\dot X_2}\bigg|_{\tau \rightarrow \infty}+O(r_s^3).  \nonumber
\end{eqnarray}
In the system under consideration, $\dot X_0=1$, $\dot Y_0=0$ and $\dot X_1 \rightarrow 0$ as $\tau \rightarrow \pm \infty $ so that
\begin{eqnarray}
\Delta\Phi&=&2(r_s\dot Y_1+r_s^2\dot Y_2)|_{\tau \rightarrow \infty}+O(r_s^3) \nonumber \\
&=&\frac{2r_s}{C_4}\bigg[1+\frac{r_s}{2C_4}+\frac{15\pi}{32}\frac{r_s}{C_4}\bigg]+O(r_s^3) \nonumber \\
&=&\frac{2r_s}{r_0}\bigg[1+\frac{r_s}{2r_0}+\frac{15\pi}{32}\frac{r_s}{r_0}\bigg]+O(r_s^3) \nonumber \\
&=&\frac{2r_s}{b}\bigg[1+\frac{15\pi}{32}\frac{r_s}{b}\bigg]+O(r_s^3) \nonumber
\end{eqnarray}
where $b=\sqrt{K_0+r_s K_1}+O(r_s^2)=C_4+r_s C_4/2+O(r_s^2)$ is the impact parameter. This deflection to second order is found to be in complete agreement with that calculated by \citet{fish}.

\subsection{Application: Travel Time Delay - second order approximation}
\begin{figure}
\begin{flushleft}
\vspace{1cm}
\caption{The top picture shows Shapiro delay ("the delay") as calculated using Gaussian Quadrature. The Middle picture shows the difference between the delay and the delay calculated using the first order approximation, and the lower picture shows the difference between the delay and that calculated using the second order approach. The vertical scale is in micro-seconds, the horizontal scale is in astronomical units (AU).}
\includegraphics[width=\textwidth]{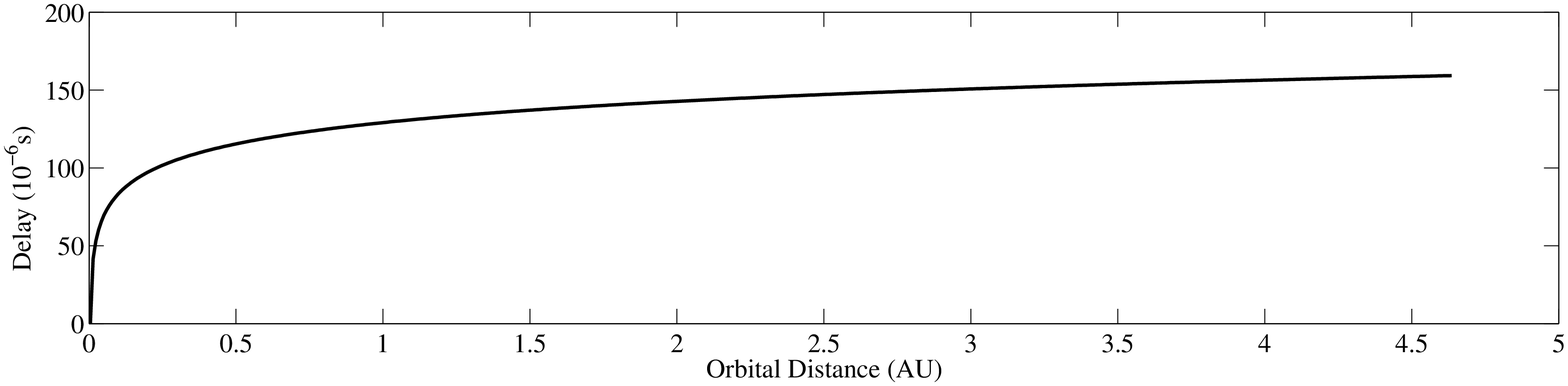} \\
\includegraphics[width=\textwidth]{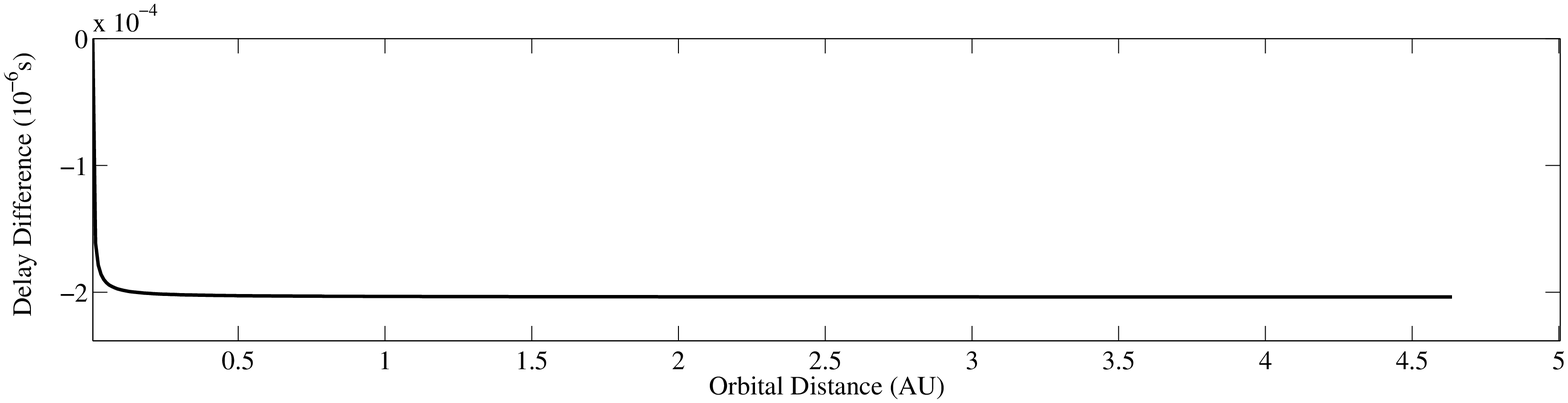} \\
\includegraphics[width=\textwidth]{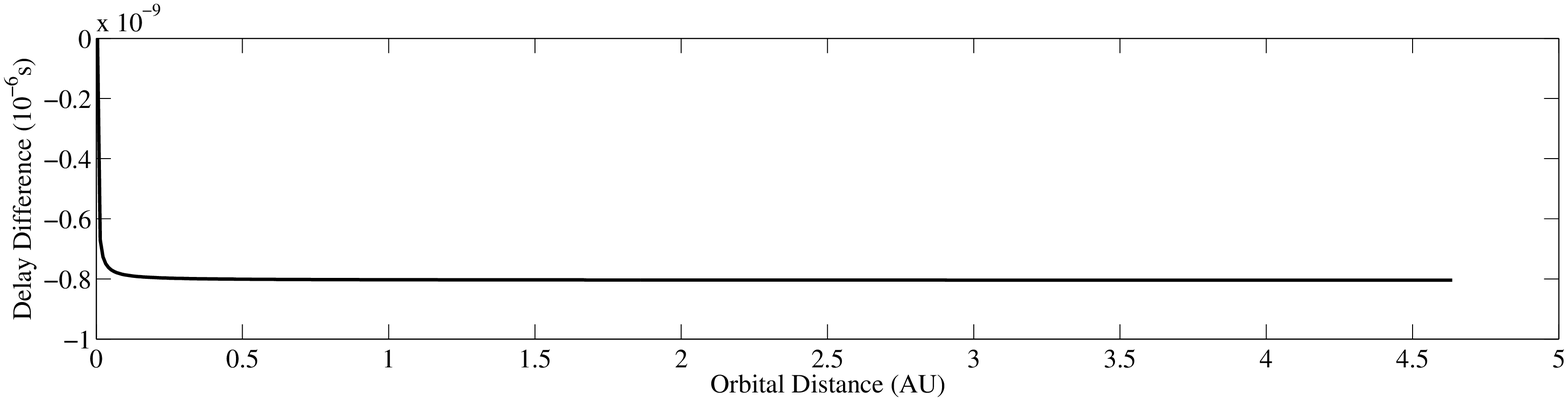}
\label{delayfig1}
\end{flushleft}
\end{figure}

As for the first-order delay calculation, we can calculate the time for the ray to go from $r_0$ to $r_f$ by solving
\begin{eqnarray}
r_f^2&=&x_f^2+y_f^2 \nonumber \\
&=& X_0^2+Y_0^2+2r_s(X_0 X_1+Y_0 Y_1)+r_s^2(X_1^2+Y_1^2+2X_0 X_2+2Y_0 Y_2) \nonumber
\label{delaycalc2}
\end{eqnarray}
for the time parameter $\tau_f$ at the final point. The initial point allows us to calculate the second order constants as $C_{12}=-1/(2r_0^2)$, $C_{22}=0$, $C_{32}=0$, $C_{42}=-9/(16r_0)$. Solving for $\tau_f$ as in Section \ref{delaysec1}, but including terms to second order in $r_s$ gives
\begin{eqnarray}
\tau_f=\sqrt{r_f^2-r_0^2}+\frac{r_s}{2}\sqrt{\frac{r_f-r_0}{r_f+r_0}}+\frac{3 r_s^2}{8 r_0} \arctan \frac{\sqrt{r_f^2-r_0^2}}{r_0}-\frac{r_s^2}{8(r_f+r_0)}\sqrt{\frac{r_f-r_0}{r_f+r_0}}+O(r_s^3).  \label{tau2}
\end{eqnarray}
Converting from $\tau$ to $t$ by integrating equation (\ref{tdot}), but this time solved to second order, results in
\begin{eqnarray}
t=\tau+r_s \ln \frac{\tau+\sqrt{\tau^2+r_0^2}}{r_0}+\frac{r_s^2}{2r_0} \bigg(3\arctan\frac{\tau}{r_0}-\frac{\tau}{\sqrt{\tau^2+r_0^2}}\bigg)+O(r_s^3).
\end{eqnarray}
This allows the second order approximation of travel time delay to be written as
\begin{eqnarray}
\Delta T&=&\frac{r_s}{2}\sqrt{\frac{r_f-r_0}{r_f+r_0}}+r_s \ln{\frac{r_f+\sqrt{r_f^2-r_0^2}}{r_0}}+r_s^2\bigg(\frac{15}{8 r_0} \arctan \frac{ \sqrt{r_f^2-r_0^2}}{r_0}-\sqrt{ \frac{r_f-r_0}{r_f+r_0}}\bigg(\frac{1}{2 r_0}+\frac{1}{8(r_f+r_0)}\bigg)\bigg).
\end{eqnarray}
In order to check this result, we may compare it to the delay ($\Delta t$) calculated numerically to high precision using Gaussian quadrature with the formula given by \citet{wfj}. For a ray starting at earth orbit, grazing the sun ($r_0=696000$km and $r_s=2.95$km) and reaching earth-orbit again, the travel time delay is calculated accurately for a range of orbital distances. In Fig. \ref{delayfig1} the delay is shown, along with the residuals from the first order and second order approaches. While the first order approximation has a relative error (that is, $(\Delta t-\Delta T)/\Delta t)$ of approximately $r_s/r_0\approx10^{-6}$, the second order approximation has a relative error of approximately $(r_s/r_0)^2\approx10^{-11}$. Distance is shown in astronomical units (`AU'), and time in micro-seconds (`$\mu$s').

\section{Rotating lens}
\begin{figure}
\vspace{1cm}
\caption{Caustic patterns due to central mass and single planet, using forward numerical integration of the full equations (\ref{schw1}). The top figure has a non-rotating central body, whereas in the bottom diagram the central body is rotating maximally (that is, with $a=r_s/2$). The light source is located on the $x$-axis at $-3 \times 10^{-6}$. The primary mass is at the origin with Schwarzschild radius $r_s=9.9 \times 10^{-7}$, and a planet is located on the $z$-axis at $3.3 \times 10^{-6}$ having $r_s=10^{-8}$. The observer's plane is located at $x=8000$.}
\includegraphics[width=15cm]{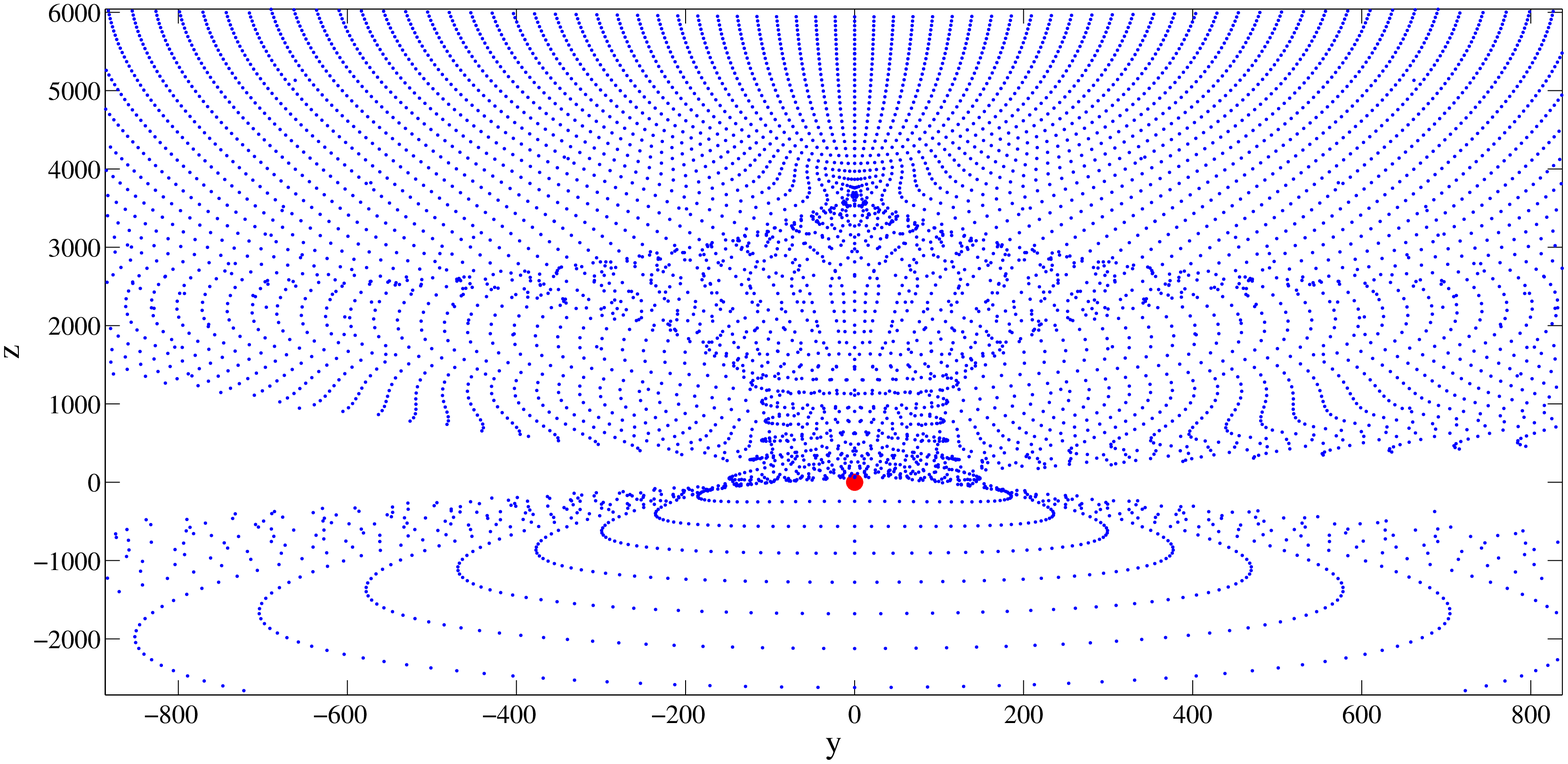} \\
\includegraphics[width=15cm]{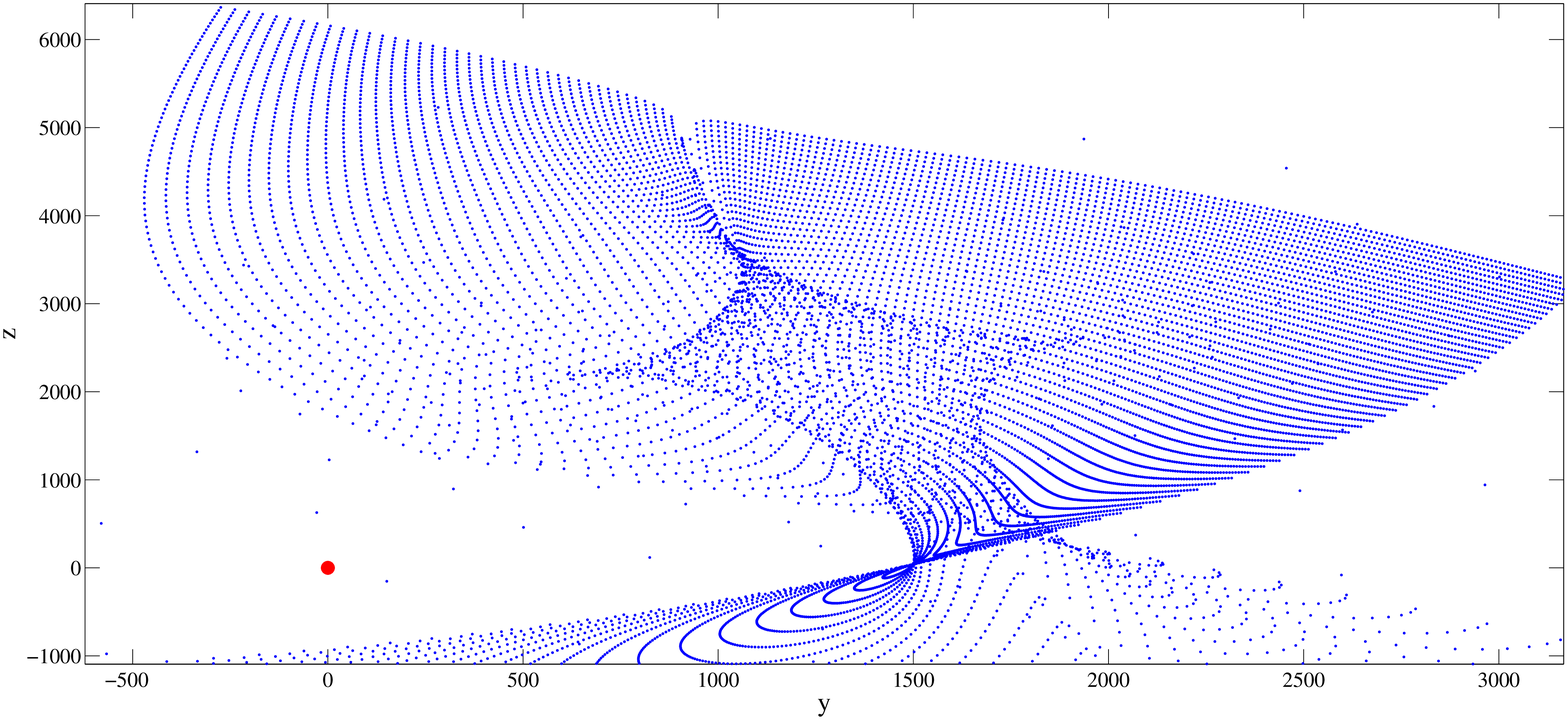}
\label{kerr1}
\end{figure}

Having explored the Schwarzschild solution in the Cartesian co-ordinate system, we are ready to move on to the rotating (Kerr) case. We may start by adding the rotational terms of the acceleration equations (\ref{xyzdd2a}) which are given in the appendix. These equations can be solved numerically using forward integration to produce a magnification map at the plane containing the observer. As the rotational terms are at second order and greater, the light rays must pass very close to the massive object to make a noticeable change to the trajectory. This is illustrated here by placing the light source close behind the massive lens. In order to observe the change in the pattern, the light source and planet have been placed approximately $3 r_s$ away from the black hole, which is clearly not a tenable position for any massive object, but is chosen only to highlight the effect of rotation on the caustic pattern. The top picture in Fig. \ref{kerr1} shows the normal diamond caustic without rotation as described by \citet{wam}, and calculated here using the numerical procedure described in \citet{wfj}. This was generated using almost 15,000 simulated light rays in a numerical integration of equation (\ref{schw1}). The lower figure uses the same procedure, but with the addition of the rotational terms. While the diamond caustic pattern is still recognizable, it has clearly undergone a twisting, with the bottom of the shape pushed further over to the right side of the diagram.

\subsection{Second order Kerr expansion}
For a black hole, physically sensible values for the rotational constant $a$ lie between $-r_s/2$ and $+r_s/2$. Therefore it is reasonable to consider $a$ to be of order $r_s$, that is $a=\alpha r_s$ where $\alpha$ is a constant between $-1/2$ and $1/2$. In the appendix, equation (\ref{xyzdd2a}) has been approximated to second order, resulting in equations (\ref{2ndorderfull}). Expanding the first of these equations using the expansions (\ref{xyz1}), yields
\begin{eqnarray}
\ddot X_0+r_s \ddot X_1+r_s^2 \ddot X_2&=&\frac{-3 r_s (X_0+r_s X_1) K_0+r_s K_0)}{2 (R_0+r_s R_1)^5}\nonumber \\
&+&r_s a \bigg( \frac{\dot Y_0}{R_0^3}+3(Y_0 \dot Z_0-Z_0 \dot Y_0)\frac{Z_0}{R_0^5}+\frac{Y_0}{R_0^4}+2\dot Y_0 \frac{\dot R_0}{R_0^3}-4Y_0\frac{\dot R_0^2}{R_0^4} \bigg) \nonumber \\
&+&a^2 \frac{2\dot X_0 Z_0}{R_0^5}(2Z_0 \dot R_0-R_0\dot Z_0) + O(r_s^3).\nonumber
\end{eqnarray}

In these equations, it can be seen that the first term on the right hand side is the Schwarzschild acceleration discussed in some detail in Section \ref{secschw}, which we have already integrated to obtain second-order path equations. It therefore remains to integrate the remaining two terms and to add them to the second-order Schwarzschild solution. The integration is straightforward, and following the same procedure for $y$ and $z$, we arrive at the second-order path equations.
\begin{eqnarray}
x&=& C_1 \tau+C_2+r_s X_1+r_s^2 X_2+r_s a \bigg(L_{x0} F_{RS} +\frac{C_3 R_0}{K_0} - \frac{Y_0}{2 R_0^2}\bigg)-a^2 C_1 F_{A} + O(r_s^3)\nonumber \\
y&=& C_3 \tau+C_4+r_s Y_1+r_s^2 Y_2+r_s a \bigg(L_{y0} F_{RS} -\frac{C_1 R_0}{K_0} + \frac{X_0}{2 R_0^2}\bigg)-a^2 C_3 F_{A} + O(r_s^3)\nonumber \\
z&=& C_5 \tau+C_6+r_s Z_1+r_s^2 Z_2+r_s a L_{z0} F_{RS} -a^2 C_5 F_{A} + O(r_s^3)\nonumber
\end{eqnarray}
in which $X_1, Y_1, Z_1$ and $X_2, Y_2, Z_2$ are described in Sections \ref{secschw1} and \ref{secschw2}. The remaining terms are
\begin{eqnarray}
F_{RS}&=&2 R_0 \frac{C_6 - B C_5}{K_0^2} - \frac{Z_0}{R_0 K_0} \nonumber \\
F_A&=&\frac{Q_0 (\tau + B)-2 C_5 K_0 Z_0}{2 R_0^2 K_0} + \frac{Q_0}{2 K_0^{3/2}} \arctan \frac{\tau+b}{\sqrt{K_0}} \nonumber \\
L_{x0}&=&C_4 C_5-C_3 C_6 \nonumber \\
L_{y0}&=&C_2 C_5-C_1 C_6 \nonumber \\
L_{z0}&=&C_2 C_3-C_1 C_4 \nonumber \\
Q_{0}&=&L_{x0}^2+L_{y0}^2. \nonumber
\end{eqnarray}
In order to estimate travel-time delays, as above we write co-ordinate time $t$ as a function of $\tau$. Expanding $\dot t$ to second order in $r_s$, and integrating yields
\begin{eqnarray}
t= \tau+r_s \log \bigg(\frac{\tau+B+R_0}{B+\sqrt{C}}\bigg)+\frac{\frac{3}{2}r_s^2+a^2}{\sqrt{K_0}}\arctan\frac{\tau\sqrt{K_0}}{B \tau+C}-r_s \tau \frac{r_s (B_R B - C_R)+ a L_{z0}}{K_0 R_0}+O(r_s^3). \nonumber
\end{eqnarray}
Again, the constants of integration have been determined by setting $t=0$ when $\tau=0$.
\subsection{Second order expansion - equatorial case}\label{sec42}
It is clear that in the equatorial case ($z=\dot z=0$, which also means that $C_5=C_6=Q_0=L_{x0}=L_{y0}=0$ and $K_0=L_{z0}^2$) the above equations simplify to
\begin{eqnarray}
x&=& C_1 \tau+C_2+r_s X_1+r_s^2 X_2+r_s a \bigg(\frac{C_3 R_0}{L_{z0}^2} - \frac{Y_0}{2 R_0^2}\bigg) + O(r_s^3)\nonumber \\
y&=& C_3 \tau+C_4+r_s Y_1+r_s^2 Y_2-r_s a \bigg(\frac{C_1 R_0}{L_{z0}^2} - \frac{X_0}{2 R_0^2}\bigg )+ O(r_s^3)\nonumber
\end{eqnarray}
Interestingly, while terms involving $r_s^2$, $r_s a$ and $a^2$ are all of second order in the expansion parameter, and terms with coefficients $r_s^2$ and $r_s a$ appear in these equatorial equations, there are no such terms with coefficient $a^2$.
\subsection{Application: Total Deflection angle - second order equatorial Kerr approximation}
We can now add the second order term due to rotation to the earlier total deflection angle calculation. As for the earlier scenario (see Fig. \ref{fig2}), $C_1=1$ and $C_3=0$ so that as $\tau$ goes to $\pm \infty$, $R_0 \rightarrow \infty$, and so $\dot y \rightarrow r_s \dot{Y_1}+r_s^2 \dot{Y_2}-r_s a \tau/(R_0 L_{z0}^2)$. Then the deflection becomes
\begin{eqnarray}
\Delta\Phi&=&2\bigg(r_s\dot Y_1+r_s^2\dot Y_2-r_s a \frac{1}{L_{z0}^2}\bigg)\bigg|_{\tau \rightarrow \infty}+O(r_s^3) \nonumber \\
&=&\frac{2r_s}{r_0}\bigg(1+\frac{r_s}{2r_0}+\frac{15\pi}{32}\frac{r_s}{r_0}-\frac{a}{r_0}\bigg)+O(r_s^3) \nonumber \\
&=&\frac{2r_s}{b}\bigg(1+\frac{15\pi}{32}\frac{r_s}{b}-\frac{a}{b}\bigg)+O(r_s^3). \nonumber
\end{eqnarray}
This deflection is found to be in complete agreement with that calculated by \citet{ede}.

\subsection{Application: Travel Time Delay - second order equatorial Kerr approximation}
As before, but including the rotational components of $x$ and $y$, the travel time for the ray to go from $r_0$ to $r_f$ can be calculated by solving 
\begin{eqnarray}
r_f^2&=&x_f^2+y_f^2 \nonumber \\
&=& X_0^2+Y_0^2+2r_s(X_0 X_1+Y_0 Y_1)+r_s^2(X_1^2+Y_1^2+2X_0 X_2+2Y_0 Y_2)+r_s a \bigg(X_0(\frac{C_3 R_0}{L_{z0}^2} - \frac{Y_0}{2 R_0^2})-Y_0(\frac{C_1 R_0}{L_{z0}^2} - \frac{X_0}{2 R_0^2})\bigg)+O(r_s^3) \nonumber \\
&=& X_0^2+Y_0^2+2r_s(X_0 X_1+Y_0 Y_1)+r_s^2(X_1^2+Y_1^2+2X_0 X_2+2Y_0 Y_2)+r_s a \frac{R_0}{L_{z0}}+O(r_s^3)
\label{delaycalc2a}
\end{eqnarray}
for the overall time $\tau_f$. As previously, the initial point allows us to calculate the second order constants. With rotation these constants become $C_{12}=-1/(2r_0^2)$, $C_{22}=a/(2r_s r_0)$, $C_{32}=-a/(2r_s r_0^2)$, $C_{42}=a/(r_s r_0)-9/(16r_0)$. The solution for $\tau_f$ now includes a rotational term (dependent on $a$), and becomes
\begin{eqnarray}
\tau_f=\sqrt{r_f^2-r_0^2}+\frac{r_s}{2}\sqrt{\frac{r_f-r_0}{r_f+r_0}}+\frac{3 r_s^2}{8 r_0} \arctan \frac{\sqrt{r_f^2-r_0^2}}{r_0}-\frac{r_s^2}{8(r_f+r_0)}\sqrt{\frac{r_f-r_0}{r_f+r_0}}+\frac{r_s a}{r_0}\sqrt{\frac{r_f-r_0}{r_f+r_0}}+O(r_s^3).  \nonumber
\end{eqnarray}
The conversion from $\tau$ to $t$ also now includes rotational terms
\begin{eqnarray}
t=\tau+r_s \ln \frac{\tau+\sqrt{\tau^2+r_0^2}}{r_0}+\frac{3r_s^2+2a^2}{2r_0} \arctan\frac{\tau}{r_0}+r_s \tau\frac{2a-r_s}{r_0\sqrt{\tau^2+r_0^2}}+O(r_s^3). \nonumber
\end{eqnarray}
This allows us to write the second order approximation of travel time delay as
\begin{eqnarray}
\Delta T&=&\frac{r_s}{2}\sqrt{\frac{r_f-r_0}{r_f+r_0}}+r_s \ln{\frac{r_f+\sqrt{r_f^2-r_0^2}}{r_0}}+\bigg(\frac{15 r_s^2}{8 r_0}+\frac{a^2}{r_0}\bigg) \arctan \frac{ \sqrt{r_f^2-r_0^2}}{r_0}+r_s\sqrt{ \frac{r_f-r_0}{r_f+r_0}}\bigg(\frac{a}{r_f}+\frac{4 a-r_s}{2 r_0}-\frac{r_s}{8(r_f+ r_0)}\bigg). \nonumber
\end{eqnarray}
This delay is the same for a ray travelling from perihelion $r_0$ to $r_f$ on the right ($t$ positive) as for a ray travelling from $r_f$ on the left ($t$ negative) to $r_0$. So the total delay for a ray passing the massive object at the origin is twice the amount $\Delta T$ stated above. It can be seen in this example, that if $a$ is positive (that is, the mass has anti-clockwise angular momentum), the motion of the particle is opposite to the frame-dragging effects, and the travel time delay is increased. Conversely, if $a$ is negative the travel time delay is decreased. \citet{dym} has calculated the delay to second order in the limit $r_f>>r_0$. The delay given here in the last equation is in agreement with Dymnikova's result in the same limit, but it also gives the second order delay for all values of $r_f$.

\section{Modelling delay for a pulsar in a binary system}

The regularity of pulses from a millisecond pulsar provides an interesting possibility for observing the effect of rotation on the travel time of the light pulses. A system such as the double pulsar binary system J0737-3039 described by \citet{bur} may provide interesting possibilities for observing the delay due to a rapidly rotating massive object. We will construct a simpler mathematical model by replacing one of the pulsars in that system with a black hole (with rotation also in the same plane as the orbit and observer) so that there is confidence in using the Kerr metric equations. Thus we consider here a binary system consisting of a millisecond pulsar and a rotating black hole with the orbital plane aligned so that the observer and the two bodies are within the same plane. We also ignore any atmospheric or magnetospheric interference which may introduce complications in measurements in the real system mentioned above. Finally we will ignore the modulation of the pulse timing due to the spinning of the pulsar. This last effect is expected to be small for a millisecond pulsar with an orbital period of hours or days, such as we are considering.

Having designed this system with the orbital plane and the observer in the equatorial plane of the black hole, we can use the simpler two dimensional equatorial equations of section \ref{sec42} to describe the paths of light rays from the pulsar to the observer. This is for simplicity and clarity only; another arrangement using the full three dimensional equations is only slightly more difficult to describe and to code.
In order to determine the delay of pulses due to the rotating black hole, we will send light rays back from the observer past the black hole using forward integration of the equatorial equations, stopping the integration procedure when the rays meet the orbital distance of the pulsar. As the time coordinate has been reversed, note that this also entails reversing the direction of spin of the black hole (which is decided arbitrarily in this model, but should be considered when using data from a real system).

\begin{figure}
\vspace{1cm}
\caption{Rays from observer to pulsar orbit at radius of $11.6$ light-seconds, using forward integration past a maximally rotating central body ($r_s=2\times10^{-4}$). Only the final sections of the rays are shown. Rays passing very close to the mass, and experiencing large deflection, are not shown here because the images produced by such rays are extremely faint.}
\includegraphics[width=\textwidth]{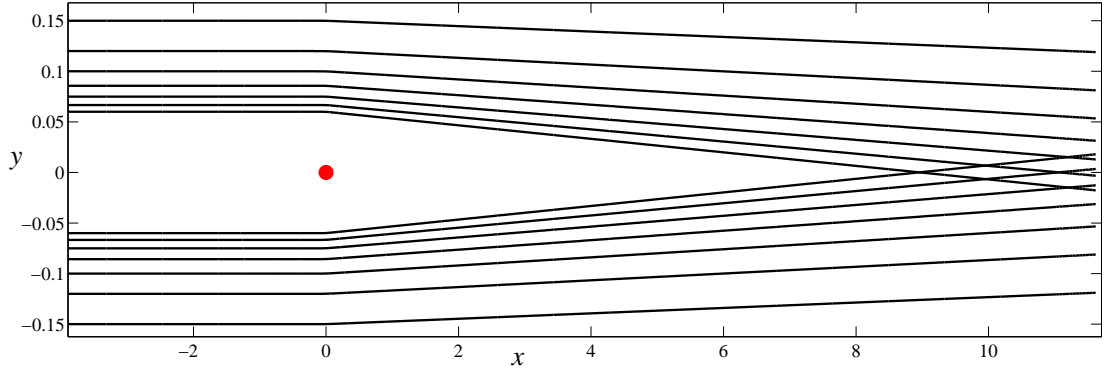}
\label{pulsar1}
\end{figure}

\begin{figure}
\vspace{1cm}
\caption{Delay due to black hole. Four curves are present for rays passing on each side of the black hole, and in the cases where the black hole is rotating (solid lines) or not rotating (dashed lines). The change in delay due to rotation is small so the curves in each pair are very close together. The lower figure shows a magnification of the central section, which allows the different delays to be seen.}
\includegraphics[width=\textwidth]{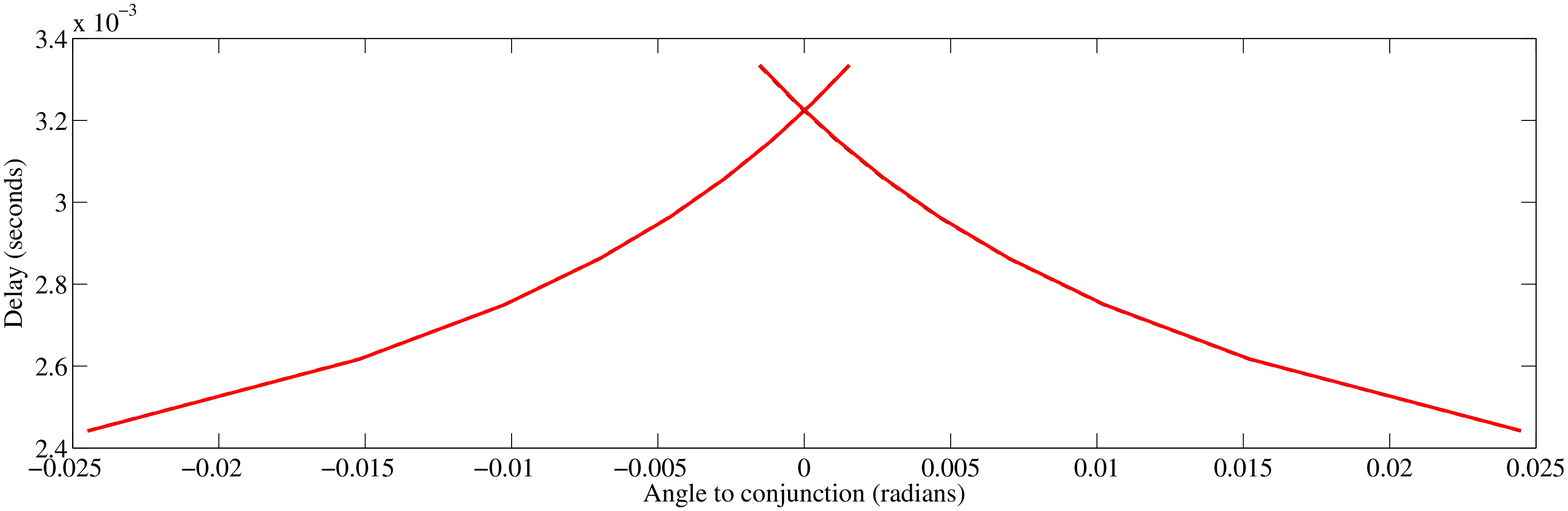}
\includegraphics[width=\textwidth]{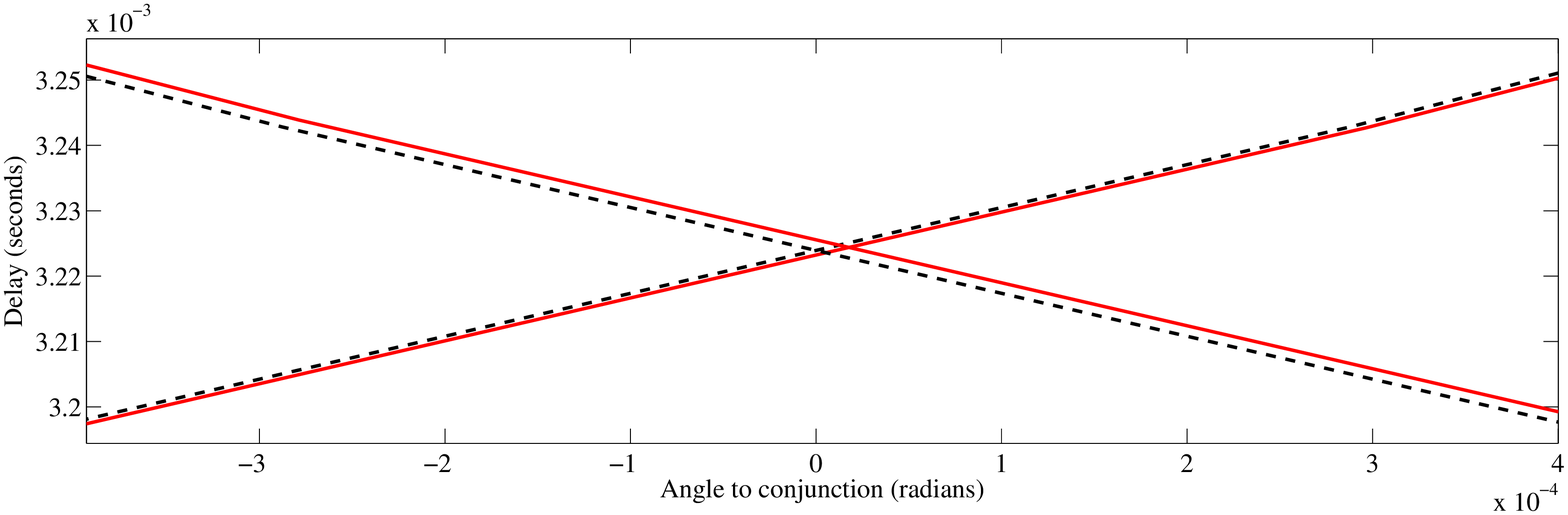}
\label{pulsar2}
\end{figure}

Figure (\ref{pulsar1}) shows the last section of rays as they reach the circular orbit of the pulsar. Due to the large difference between the vertical and horizontal scales, and because only a small section of the orbit is shown, the endpoints of the rays appear to be in a line, but they do in fact form a circular arc. All distances shown are in light-seconds and times for the delays are in seconds. The Schwarzschild radius ($r_s$) of the black hole is $2 \times 10^{-4}$ light-seconds, equivalent to approximately $20$ solar masses. The delay increases in an almost linear relationship with mass of the lens, so that a black hole of $10$ solar masses would have approximately half the delay times as those shown in Fig. \ref{pulsar2}. In a different study of travel time delay in a binary pulsar system, \citet{lag} note that in order for the binary system to have sufficient longevity for a reasonable chance of observation, there are limitations on the proximity of the pulsar to the black hole, with approximately $5$ solar radii being near optimum compromise between longevity and magnitude of the delay effect. We therefore place the circular orbit at $11.6$ light-seconds, approximately $5$ solar radii. This orbit induces a delay term in the straight-line time from $-11.6$ seconds when the pulsar is closest to the observer, to $+11.6$ seconds when it is furthest from the observer. However, in the present study, we are interested in the additional asymmetric delay due to rotation. The orbital delay is symmetric about the point of superior conjunction, or occultation, which occurs when the lensing body is directly between the pulsar and the observer, and is the point of interest for the present study. For this reason, the orbital delay will be ignored here.

Along with the equatorial equations, we also integrate Equation (\ref{tdot}) to keep track of the time coordinate. Comparing the time taken with the time light would take to travel in a straight line from pulsar to observer without any lensing object gives the delay, shown in Fig. \ref{pulsar2}. The solid curve represents the delay due to a rotating black hole, and was calculated using forward integration of the equatorial equations (\ref{eqnequat1}) given in the appendix. This delay is plotted against the angle from superior conjunction. That is, a zero angle represents the system with the black hole in between the pulsar and observer, while all three are in a line. The dashed line represents the delay due to a non-rotating black hole, calculated using the Schwarzschild acceleration, equation (\ref{schw1}). The part of the delay due to rotation is small, and so to highlight the difference between the delays shown, a magnified portion near the intersection is shown in the lower figure.

\begin{figure}
\vspace{1cm}
\caption{Difference in travel-time delay between rotating and non-rotating cases, plotted against orbital position of the pulsar. Each curve corresponds to rays passing the black hole on either the left (opposed to the black hole's rotation) or right side (aligned to the rotation).}
\includegraphics[width=\textwidth]{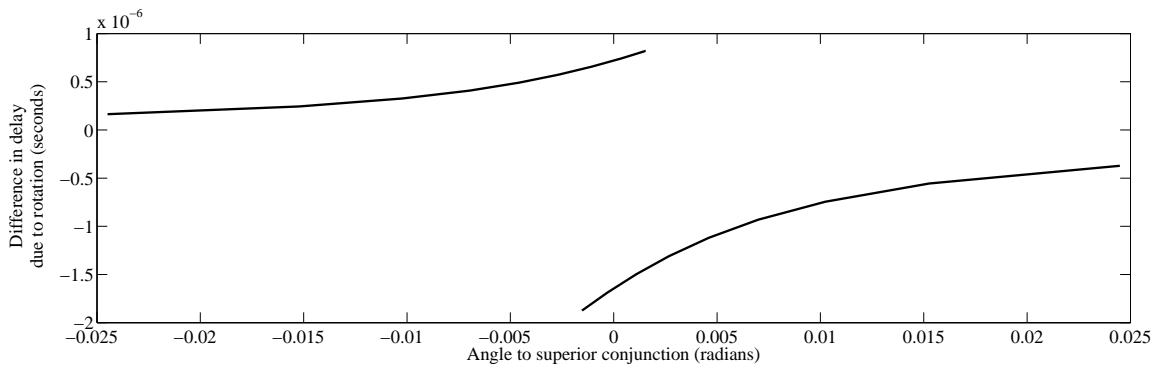}
\label{pulsar3}
\end{figure}

Finally the delay due to maximal rotation of the black hole is subtracted from the delay due to a black hole of the same mass without rotation. The difference between these delays gives the delay due solely to rotation, which provides a small asymmetry in the delay curve in the rotational case. This difference is shown in Figure (\ref{pulsar3}). The upper and lower lines represent the delay difference for the two different images of the source, one passing to the left of the mass, the other to the right. One of these images is dominant prior to conjunction, and the other image becomes dominant after conjunction.

The delays due to lensing in this system are of the order of $10^{-3}$ seconds, while the asymmetric part of the delay that is due to rotation is of the order of $10^{-6}$ seconds. Assuming other effects on travel time can be adequately accounted for, these rotational delays may be measurable, and could possibly be used to estimate the angular momentum of the lens.

\section{Conclusion and Discussion}

In this paper, we have presented Cartesian acceleration components for photons using the Kerr metric. While these components are somewhat complicated, they allow easy modelling of systems in Cartesian coordinates, with the advantage that the components are normally very small. This allows for rapid numerical integration, and a new result (the caustic shape due to a binary system with rotating mass) has been presented. In order to approximate the light paths near a rotating black hole, we have built up the approximations in stages, beginning with the zeroth order and first order expansions, followed by the second order Schwarzschild expansion and finally the second order Kerr expansion. At each point, the versatility of this approach has been demonstrated by the ease of calculating deflection and travel time delay, which are found to match previously calculated amounts. In addition, a new formula for delay due to spinning black holes was presented. It may prove possible in some practical astrophysical circumstances to measure directly the delay due to rotation of the lensing object, and so to infer the angular momentum, using the formulae presented here. This awaits future experimental observation.

\section{Appendix}
In this appendix, the Cartesian acceleration components are written out in full. The dot denotes differentiation by the chosen parameter $\tau$. These equations are followed by their second order approximations. Next the $x$ and $y$ acceleration components for the equatorial special case are presented, followed by their second order approximations.
\subsection{Three-Dimensional Acceleration Components}

Expanding and simplifying equation (\ref{xdd}) for $\ddot x$ and following the same procedure for $\ddot y$ and $\ddot z$, we can derive the following forms of the acceleration components:

\begin{eqnarray}
\ddot x&=&-\frac{3 r_s x}{2 r^3} \frac{Q+(L-a)^2}{r^2+a^2}-\frac{a^2 \dot r \dot x}{r}\bigg(\frac{1}{a^2+r^2}+\frac{z^4}{r^2 (r^2-z^2)^2}\bigg)+\frac{a x}{r^2+a^2} G_1+\frac{a y}{\Delta} G_2 \nonumber \\
\ddot y&=&-\frac{3 r_s y}{2 r^3} \frac{Q+(L-a)^2}{r^2+a^2}-\frac{a^2 \dot r \dot y}{r}\bigg(\frac{1}{a^2+r^2}+\frac{z^4}{r^2 (r^2-z^2)^2}\bigg)+\frac{a y}{r^2+a^2} G_1-\frac{a x}{\Delta} G_2 \nonumber \\
\ddot z&=&-\frac{3 r_s z (Q+(L-a)^2)}{2 r^5}+\frac{2 a^2 z (Q-z^2)}{r^6}.
\label{xyzdd2a}
\end{eqnarray}
The functions $G_1$ and $G_2$ are functions of $x,y,z,\dot x, \dot y, \dot z$ and are of order $r_s$. The axi-symmetry of the system may be seen in the signs in the equations for $\ddot x$ and $\ddot y$. That is, the equations are invariant under a rotation of the system about the $z$-axis such as given by the transformations $x \rightarrow y, y \rightarrow -x$. The functions $G_1$ and $G_2$ are expressed below:

\begin{eqnarray}
G_1&=& a \dot r^2 \bigg(\frac{2}{a^2+r^2}+\frac{z^4 (a^2 z^2+r^4)}{r^4 (r^2-z^2)^3}-\frac{z^2}{r^2 (r^2-z^2)}\bigg)+\frac{a z \dot r \dot z (r^6-a^2 z^4-2 r^4 z^2)}{r^3 (r^2-z^2)^3} \nonumber \\
&-&\frac{a}{r^6} (2 a^2 z^2+r^4+2 r^2 z^2)+\frac{a L^2}{r^4} \bigg(\frac{r^2+z^2}{r^2-z^2}+\frac{2 r^3 r_s}{\Delta  (r^2-z^2)}-\frac{a^2 (a^2+r^2)}{\Delta ^2}\bigg) \nonumber \\
&-&\frac{r_s}{r \Delta}\big(a^2+r^2-r^2 \dot r\big)  \bigg[2 L \bigg(\frac{1}{r^2-z^2}-\frac{a^2}{\Delta  r^2}\bigg)+a\frac{r_s}{r \Delta}\bigg(1-\frac{r^2 \dot r}{a^2+r^2}\bigg)\bigg]+\frac{a Q}{r^4} \nonumber
\end{eqnarray}

\begin{eqnarray}
G_2 &=& r_s \frac{2 r^2 (Q+L^2-a^2)+4 a^2 Q-3 r r_s (Q+(L-a)^2)}{2 r^4 (a^2+r^2)}+\frac{2 z (z \dot r-r \dot z)}{r^3 (r^2-z^2)} \bigg(\frac{r^3 r_s \dot r}{a^2+r^2}+a L-r r_s \bigg) \nonumber \\
&+& \frac{r_s \dot r^2}{a^2+r^2} (\frac{a^2-r^2}{\Delta }+\frac{2 a^2}{a^2+r^2})-\frac{\dot r}{r^2 \Delta }\big(r_s (a^2-r^2)+a L (2 r-r_s)\big) \nonumber \\
&+& \frac{a \dot r}{r^2} \bigg(\frac{r^2-2 z^2}{(r^2-z^2)^2}-\frac{a^2}{r^2 (a^2+r^2)}\bigg)\bigg(r_s a+\frac{\Delta  L r}{r^2-z^2}-\frac{a^2 L}{r}-\frac{a r^2 r_s \dot r}{a^2+r^2}\bigg) \nonumber
\end{eqnarray}

The constants of motion $L$ and $Q$ and also $r$ and $\Delta$ can be expressed as the following functions of $x,y,z,\dot x, \dot y, \dot z$:
\begin{eqnarray}
r &=& \sqrt{\frac{x^2+y^2+z^2-a^2+\sqrt{(x^2+y^2+z^2-a^2)^2+4 a^2 z^2} }{2}}  \nonumber \\
L &=& \frac{r(r^2-z^2)}{r^4 - r_s r^3 + a^2 z^2}\bigg( r \Delta \frac{x \dot y-y \dot x}{x^2 + y^2}-r_s a (1-\frac{r^2 \dot r}{a^2+r^2})\bigg) \nonumber  \nonumber \\
Q &=& \frac{r^2(z \dot r-r \dot z)^2+L^2 z^2}{r^2-z^2}-\frac{a^2 z^2}{r^2} \nonumber \\
\Delta &=& a^2 + r^2 - r_s r \nonumber
\end{eqnarray}

\subsection{Second Order Approximation}
In order to approximate  $\ddot x, \ddot y$ and $\ddot z$ to second order in $r_s$, it is only necessary to approximate $G_1$ and $G_2$ to first order in $r_s$, as they are multiplied by $a$, which is of order $r_s$. We first express the intermediary functions to first order in $r_s$. Ignoring terms of second order and higher in $r_s$, where $a$ is of order $r_s$, we have the following simplifications:
\begin{eqnarray}
L &=&L_z +O(r_s^2)= x \dot y-y \dot x +O(r_s^2) \nonumber \\
r &=& \sqrt{x^2+y^2+z^2} +O(r_s^2)  \nonumber \\
Q &=& L_x^2+L_y^2 +O(r_s^2) = (y \dot z-z \dot y)^2 +(x \dot z-z \dot x)^2+O(r_s^2) \nonumber \\
\Delta &=& r^2 - r_s r +O(r_s^2) \nonumber
\end{eqnarray}
So that $G_1$ and $G_2$ can be approximated by:

\begin{eqnarray}
G_1&=& a \dot r^2 \bigg(\frac{2}{r^2}+\frac{z^4}{(r^2-z^2)^3}-\frac{z^2}{r^2 (r^2-z^2)}\bigg)+\frac{a r \dot r z \dot z (r^2-2 z^2)}{(r^2-z^2)^3}+\frac{2 r_s L}{r}\frac{\dot r-1}{r^2-z^2}+\frac{a}{r^4}\bigg(Q+L^2\frac{r^2+z^2}{r^2-z^2}- r^2-2 z^2\bigg) \nonumber \\
G_2 &=&  \frac{r_s (Q+L^2)}{r^4}+\frac{2 z (z \dot r-r \dot z)}{r^3 (r^2-z^2)} \bigg(r r_s (\dot r-1)+a L\bigg)- \frac{r_s \dot r}{r^2}(\dot r-1)-2 a L\frac{\dot r}{r^3}+a L r\dot r \frac{r^2-2 z^2}{(r^2-z^2)^3} \nonumber
\end{eqnarray}

We can then write the acceleration components, $\ddot x, \ddot y, \ddot z$ as:

\begin{eqnarray}
\ddot x&=&-\frac{3 r_s x}{2 r^5} (Q+L^2-2 a L)-\frac{a^2 \dot r \dot x}{r^3}\bigg(1+\frac{z^4}{(r^2-z^2)^2}\bigg)+\frac{a x}{r^2} G_1+\frac{a y}{r^2} G_2+O(r_s^3) \nonumber \\
\ddot y&=&-\frac{3 r_s y}{2 r^5} (Q+L^2-2 a L)-\frac{a^2 \dot r \dot y}{r^3}\bigg(1+\frac{z^4}{(r^2-z^2)^2}\bigg)+\frac{a y}{r^2} G_1-\frac{a x}{r^2} G_2+O(r_s^3) \nonumber \\
\ddot z&=&-\frac{3 r_s z (Q+L^2-2 a L)}{2 r^5}+\frac{2 a^2 z (Q-z^2)}{r^6}+O(r_s^3) \nonumber
\end{eqnarray}
After some algebra, these may be written as:

\begin{eqnarray}
\ddot x&=&-\frac{3 r_s x}{2 r^5} (Q+L^2)+r_s a \bigg(\frac{\dot y}{r^3}+3(y \dot z-z \dot y)\frac{z}{r^5}+\frac{y}{r^4}+2\dot y \frac{\dot r}{r^3}-4y\frac{\dot r^2}{r^4}\bigg)-a^2\frac{2\dot x z}{r^5}(2z \dot r-r\dot z)+O(r_s^3) \nonumber \\
\ddot y&=&-\frac{3 r_s y}{2 r^5} (Q+L^2)-r_s a \bigg(\frac{\dot x}{r^3}+3(x \dot z-z \dot x)\frac{z}{r^5}+\frac{x}{r^4}+2\dot x \frac{\dot r}{r^3}-4x\frac{\dot r^2}{r^4}\bigg)-a^2\frac{2\dot y z}{r^5}(2z \dot r-r\dot z)+O(r_s^3) \nonumber \\
\ddot z&=&-\frac{3 r_s z }{2 r^5}(Q+L^2)-r_s a\frac{3 L z}{r^5}+a^2 \frac{2 z (Q-z^2)}{r^6}+O(r_s^3)
\label{2ndorderfull}
\end{eqnarray}

\subsection{Equatorial Equations}
From equation (\ref{xyzdd2a}), we can see that a particle in the $x-y$ plane, has no component of acceleration in the $z$ direction. That is, if $z=0$, then $\ddot z = 0$. If such a particle also has no velocity component in the $z$ direction it must remain within the plane. Thus there is a 2-D special case. The equations are derived by setting $z$ and $\dot z$ to zero. This also means that $Q$ is zero. Then we have the following acceleration components:

\begin{eqnarray}
\ddot x=-\frac{3 r_s x}{2 r^3} \frac{(L-a)^2}{r^2+a^2}-\frac{a^2 \dot r \dot x}{r} \frac{1}{a^2+r^2}+\frac{a x}{r^2+a^2} G_1+\frac{a y}{\Delta} G_2 \nonumber \\
\ddot y=-\frac{3 r_s y}{2 r^3} \frac{(L-a)^2}{r^2+a^2}-\frac{a^2 \dot r \dot y}{r} \frac{1}{a^2+r^2}+\frac{a y}{r^2+a^2} G_1-\frac{a x}{\Delta} G_2 \label{eqnequat1}
\end{eqnarray}
Where $G_1$ and $G_2$ are now given by:
\begin{eqnarray}
G_1&=& a \dot r^2 \bigg(\frac{2}{a^2+r^2}\bigg)-\frac{a}{r^2}+\frac{a L^2}{r^4} \bigg(1+\frac{2 r r_s}{\Delta  }-\frac{a^2 (a^2+r^2)}{\Delta ^2}\bigg) \nonumber \\
&-&\frac{r_s}{r \Delta}\big(a^2+r^2-r^2 \dot r\big)  \bigg[\frac{2 L}{r^2} \bigg(1-\frac{a^2}{\Delta}\bigg)+\frac{r_s a}{r \Delta}\bigg(1-\frac{r^2 \dot r}{a^2+r^2}\bigg)\bigg] \nonumber \\
G_2 &=&  \frac{r_s}{a^2+r^2}\bigg[\dot r^2\bigg(\frac{a^2}{a^2+r^2}+\frac{a^2-r^2}{\Delta}\bigg)+\frac{2 r (L^2-a^2)-3 r_s (L-a)^2}{2 r^3}- \frac{\dot r}{r \Delta}(a L(r-r_s)+r_s a^2)\bigg]+\frac{\dot r}{r \Delta}(r_s r-a L) \nonumber
\end{eqnarray}
In which, $r$, $\dot r$ and $L$ are given by
\begin{eqnarray}
r^2 &=&  x^2+y^2-a^2  \nonumber \\
\dot r &=&  \frac{x \dot x+y \dot y}{r} \nonumber \\
L &=& \frac{1}{r - r_s}\bigg( r \Delta \frac{x \dot y-y \dot x}{x^2 + y^2}-r_s a (1-\frac{r^2 \dot r}{a^2+r^2})\bigg) \nonumber
\end{eqnarray}

\subsection{Equatorial Case: Second Order Approximation}
Discarding terms of order higher than $r_s^2$ (where again, $a$ is of order $r_s$), we obtain the following acceleration components:
\begin{eqnarray}
\ddot x=-\frac{3 r_s x}{2 r^5} (L^2-2 a L)-\frac{a^2 \dot r \dot x}{r^3}+\frac{a x}{r^2} G_1+\frac{a y}{r^2} G_2 \nonumber \\
\ddot y=-\frac{3 r_s y}{2 r^5}(L^2-2 a L)-\frac{a^2 \dot r \dot y}{r^3}+\frac{a y}{r^2} G_1-\frac{a x}{r^2} G_2 \nonumber
\end{eqnarray}
Where $G_1$ and $G_2$ are now given by:
\begin{eqnarray}
G_1&=& a \dot r^2 \bigg(\frac{2}{r^2}\bigg)-\frac{a}{r^2}+\frac{a L^2}{r^4}-\frac{2 r_s L}{r^3}\big(1- \dot r\big) \nonumber \\
G_2 &=&  \frac{r_s \dot r}{r^2}\bigg[1-\dot r+\frac{L^2}{r^2}\bigg]-\frac{a L \dot r}{r^3} \nonumber
\end{eqnarray}
In which, $r$, $\dot r$ and $L$ are given by
\begin{eqnarray}
r^2 &=&  x^2+y^2 \nonumber \\
\dot r &=&  \frac{x \dot x+y \dot y}{r} \nonumber \\
L &=&L_z +O(r_s^2)= x \dot y-y \dot x +O(r_s^2) \nonumber
\end{eqnarray}

After some algebra, in which all the terms associated with $a^2$ cancel out, $\ddot x$ and $\ddot y$ may be written as:

\begin{eqnarray}
\ddot x&=&-\frac{3 r_s x}{2 r^5} L^2+r_s a \bigg(\frac{\dot y}{r^3}+\frac{y}{r^4}+2\dot y \frac{\dot r}{r^3}-4y\frac{\dot r^2}{r^4}\bigg)+O(r_s^3) \nonumber \\
\ddot y&=&-\frac{3 r_s y}{2 r^5} L^2-r_s a \bigg(\frac{\dot x}{r^3}+\frac{x}{r^4}+2\dot x \frac{\dot r}{r^3}-4x\frac{\dot r^2}{r^4}\bigg)+O(r_s^3) \label{xyddeq2ndb}
\end{eqnarray}

\end{document}